\documentclass[pra,showpacs,twocolumn,eqsecnum]{revtex4}
\usepackage{dcolumn}% Align table columns on decimal point
\usepackage{bm}% bold math
\usepackage{amssymb}
\usepackage{amsmath}
\usepackage{amsfonts}

\usepackage[]{graphicx}
\begin{document}
\bibliographystyle{prsty}
\title{Description of spontaneous photon emission  and local density of states in presence of a lossy polaritonic inhomogenous  medium  }
\author{ Aur\'elien Drezet $^{1}$}
\address{(1) Univ.~Grenoble Alpes, CNRS, Institut N\'{e}el, F-38000 Grenoble, France}
\begin{abstract}
We provide a description of spontaneous emission in a dispersive and dissipative linear inhomogeneous medium based on the generalized Huttner-Barnett model [Phys. Rev. A 46, 4306 (1992)]. Our discussion considers on an equal footing both the photonic and material fluctuations which are necessary to preserve unitarity of the quantum evolution.  Within this approach we justify the results obtained  in the past using the Langevin noise method that neglects the removal of photonic fluctuations. We finally discuss the concept of local density of states (LDOS) in a lossy and dispersive inhomogeneous environment that provides a basis for theoretical studies of fluorescent emitters near plasmonic and polaritonic antennas.          
\end{abstract}

 \maketitle
%%%%%%%%%%%
\section{Introduction}
In the recent years the theoretical problem of describing the coupling of single fluorescent quantum emitters with a metallic nano-particle supporting surface plasmon (SP) modes has become very urgent due to many applications  envisioned with photonic and quantum information processing technologies at the nanoscale.  In particular  the concept of local density of states (LDOS) \cite{Colas,Novotny} is central since it provides a figure a merit for quantifying the coupling of quantum emitters to plasmonic systems and it plays a central role in recent studies using near-field optical microscopes~\cite{Colas,Colas2,Chicanne,Colas3,Pham,Martin}. However, one of the main issue with plasmonic systems is that they are intrinsically dissipative and that a self-consistent quantum electrodynamics (QED) description  of plasmons, i.e. respecting rigorously the unitarity of time evolution,   involves necessarily the inclusion of additional degrees of freedom associated with fluctuating  currents and dipoles in the metal. This problem has been generally studied  using the Langevin noise method advocated originally by Gruner and Welsch~\cite{Gruner1995,Gruner1996,Yeung1996} and it has been intensively used meanwhile in QED in macroscopic media~\cite{Scheel1998,Dung1998,Dung2000,Scheel2001,Matloob1999,Matloob2004,Fermani2006,Raabe2007,Amooshahi2008,Scheelreview2008}, e.g., for calculating the coupling of quantum fluorescent emitters to fluctuating fields near plasmonic antennas~\cite{Dzotjan2010,Cano2011,Hummer2013,Chen2013,Delga2014,Hakami2014,Choquette2012,Grimsmo2013,Rousseaux2016}. The justification and consistency of the approach (which intuitively generalizes earlier fundamental results obtained by Rytov and Lifschitz in the context of Casimir force calculations~\cite{Lifshitz1956,Ginzburg,Rytov,Milonnibook,Callen1951,Rosa2010,Agarwal1975,Sipe1984,Carminati1999,Mulet,Rousseau,Henkel}) has been however the subject of some controversies in the past and a rigorous mathematical derivation based on a QED hamiltonian formalism valid for the most general inhomogeneous media  has been looked  for during years.   The main issue is that the Langevin noise approach was originally motivated by the Huttner-Barnett  Hamiltonian formalism ~\cite{Huttner1991,Huttner1992a,Huttner1992b,Huttner1992c,Matloob1995,Matloob1996,Barnett1995} which for a homogeneous dissipative and dispersive dielectric material, respecting Kramers-Kr\"{o}nig causality constraints, generalizes the historical Hopfield-Fano model for polaritons in bulk media \cite{Fano1956,Hopfield1958}. The Huttner-Barnett formalism leads directly to the results postulated by Gruner and Welsch in their seminal work and in particular it was used  in the calculation of the emission rate of fluorescent dipoles imbedded in a lossy dielectric matrix~\cite{Huttner1991,Huttner1992a,Huttner1992b,Huttner1992c,Matloob1995,Matloob1996,Barnett1995}. However, the rigorous equivalence between the two formalisms in the more general inhomogeneous situation was still lacking for years.\\
\indent Very recently, based on important calculation by  Suttorp and coworkers, several works~\cite{Wubs2001,Suttorp2004a,Suttorp2004b,Suttorp2007,Bhat2006,Judge2013,Philbin2010} proposed a mathematical justification of the Langevin noise method based on the direct generalization of the Huttner-Barnett model for inhomogeneous media. However, we showed in two publications \cite{A}, and \cite{B} that these earlier derivations overlooked the role of boundary conditions concerning the value taken by dielectric susceptibilities at spatial infinity.  We showed that to preserve the fundamental unitarity of the quantum evolution one must  necessarily  add a contribution associated with vacuum photonic fluctuation to the fluctuating source current terms calculated  by Gruner and Welsch~\cite{Gruner1995,Gruner1996,Yeung1996}. Actually, it implies that, rigorously speaking, the formalism introduced  by Gruner and Welsch ~\cite{Gruner1995,Gruner1996,Yeung1996} relies on the assumption that the surrounding dielectric environment is necessarily lossy and this, even at spatial infinity. Contrarily to a widespread belief this situation is  not generally applicable to nanoparticle antennas which are by definition spatially localized  and very often surrounded by vacuum  in the calculations.\\
\indent The main objective of the present work is to leverage on the analysis started by us in \cite{A,B} in order to describe the coupling of quantum emitters to fluctuating currents and fields. In the present paper we will consider specifically the regime of spontaneous emission and the coupling of transition dipoles to coherent laser sources (i.e., the derivation of the optical Bloch equations). We will demonstrate that the most general formalism derived in \cite{B} actually allows us to generalize the results obtained for spontaneous emission that were only rigorously derived for a dipole in vacuum. We will show that for this specific problem the dynamical equations and emission rates actually agree with previous results obtained using the Gruner and Welsch method without the assumptions made in this phenomenological approach.\\
\indent  The layout of this paper is a follows: In section II we provide a general description of Huttner-Barnett formalism~\cite{Huttner1992a,A,B} using the dual Lagrangian method given in \cite{A,B}. We include in the model the interaction between  a fluorescent molecule and the photonic and dielectric environment.  In section III we analyze the spatio-temporal evolution of the electric  field operator  and separate the contributions associated with free photons from fluctuating currents (associated with the dielectric medium and the molecule). In sect IV we study within the Wigner-Weisskopf method the spontaneous emission of a two-level atom in a general lossy and dissipative inhomogeneous environment. We conclude with a discussion in section V about LDOS and photonic wave functions associated with spontaneously emitted photons and we derive the dynamics associated with the optical Bloch equations.                                    
%%%%%%%%%%%%%%%%
\section{The dynamical equations of a transition dipole in a fluctuating photonic and dielectric environment }
We start with the dual-Lagrangian density for the coupled system 
\begin{eqnarray}
\mathcal{L}=\frac{\mathbf{B}^2-\mathbf{D}^2}{2}+ \mathbf{F}\cdot\boldsymbol{\nabla}\times\mathbf{P} -\frac{\mathbf{P}^2}{2}+\mathcal{L}_M+\mathcal{L}_{\Psi_a}\label{1bb}
\end{eqnarray}
where by definition $\mathbf{B}(\mathbf{x},t)=\frac{1}{c}\partial_t\mathbf{F}(\mathbf{x},t)$,  and $\mathbf{D}(\mathbf{x},t)=\boldsymbol{\nabla}\times\mathbf{F}(\mathbf{x},t)$ and where we use the Coulomb gauge constraint $\boldsymbol{\nabla}\cdot\mathbf{F}(\mathbf{x},t)=0$ for the electric potential. The electric potential $\mathbf{F}(\mathbf{x},t)$ is the dual of the usual magnetic potential $\mathbf{A}(\mathbf{x},t)$ and we showed that it is a specifically adapted choice for quantization of electromagnetic problems involving dipole densities $\mathbf{P}(\mathbf{x},t)$ \cite{A,B}. In this description the material part associated with the inhomogeneous dielectric medium reads 
\begin{eqnarray}
\mathcal{L}_M=\int_{0}^{+\infty}d\omega\frac{(\partial_t\mathbf{X}_\omega)^2-\omega^2\mathbf{X}_\omega^2}{2}.\label{2}
\end{eqnarray}  and corresponds to the Huttner-Barnett model \cite{Huttner1992a,A,B}. In this model the field $\mathbf{X}_\omega(\mathbf{x},t)$ completely characterizes  the fully causal dielectric environment satisfying Kramers-Kr\"{o}nig relations~\cite{Huttner1992a,A,B}.\\
\indent The contribution~\cite{Power} 
\begin{eqnarray}
\mathcal{L}_{\Psi_a}=i\hbar\Psi_a^\ast\partial_t\Psi_a-V\Psi_a^\ast\Psi_a\nonumber\\-\frac{\hbar^2}{2M}\boldsymbol{\nabla}\Psi_a^\ast \cdot\boldsymbol{\nabla}\Psi_a\label{psi}
\end{eqnarray} is associated with the Schr\''{o}dinger matter fields $\Psi_a(\mathbf{x},t)$ and $\Psi_a^\ast(\mathbf{x},t)$ which describe the atomic dipole of mass $M$ in the external potential $V(\mathbf{x})$. In the second quantization formalism we expand the atomic wave function operator as $\Psi_a(\mathbf{x},t)=\sum_m b_m(t)\psi_m(\mathbf{x})$ where $\psi_m(\mathbf{x})$ (labeled by $m$) are energy eigenstates of the time independent Schrodinger equation $E_m\psi_m=(-\frac{\hbar^2\boldsymbol{\nabla}^2}{2m}+V)\psi_m$ defined for the energy $E_m$ and forming a complete orthogonal set such that $\int d^3\mathbf{x}\psi_m^\ast(\mathbf{x})\psi_n(\mathbf{x})=\delta_{n,m}$. The usual second quantization procedure leads to the (fermionic) anti-commutators
 \begin{eqnarray}
\{b_n(t),b_m^\dagger (t)\}=\delta_{n,m},\label{rififi}
\end{eqnarray}  and
$\{b_n(t),b_m^\dagger (t)\}=0=\{b_n(t)^\dagger,b_m^\dagger (t)\}$. This is clearly consistent with the usual canonical quantization procedure  since the Lagrangian function $L_{\Psi_a}=\int d^3\mathbf{x}\mathcal{L}_{\Psi_a}(\mathbf{x},t)=i\hbar\sum_{m}b_m^\dagger\dot{b}_m-\sum_{m}E_m b_n^\dagger,b_m^\dagger$ implies the canonical momenta:
\begin{eqnarray}
\Pi_{b_m}=\frac{\partial L_{\Psi_a}}{\partial \dot{b}_m}=i\hbar b_n(t)^\dagger,
\Pi_{b_m^\dagger}=\frac{\partial L_{\Psi_a}}{\partial \dot{b}_m^\dagger}=0
\end{eqnarray} which together with the canonical anti-commutation rules $\{b_m,\Pi_{b_n}\}=i\hbar\delta_{n,m}$, etc... imply Eq.~\ref{rififi}.\\
\indent Moreover, in this model the dipole density $\mathbf{P}(\mathbf{x},t)$ includes contributions from  the surrounding dielectric medium $\mathbf{P}_{\textrm{diel.}}(\mathbf{x},t)$ and of the Schrodinger field $\mathbf{P}_\Psi(\mathbf{x},t)$. For the dielectric part we use in agreement with Huttner and Barnett (see \cite{Huttner1992a,A,B}):
\begin{eqnarray}
\mathbf{P}_{\textrm{diel.}}(\mathbf{x},t)=\int_{0}^{+\infty}d\omega\sqrt{\frac{2\sigma_\omega(\mathbf{x})}{\pi}}\mathbf{X}_\omega(\mathbf{x},t).\label{5new}
\end{eqnarray}  
For the Schrodinger field we here model the dipole density fluid using the formula: 
\begin{eqnarray}
\mathbf{P}_\Psi(\mathbf{x},t)\simeq e\int d^3\boldsymbol{\xi}\Psi_a^\ast(\boldsymbol{\xi},t)\boldsymbol{\xi}\Psi_a(\boldsymbol{\xi},t)\Delta(\mathbf{x}-\mathbf{x}_0)\nonumber\\
=\sum_{m,n}b_m^\dagger(t) b_n(t)\boldsymbol{\mu}_{m,n}\Delta(\mathbf{x}-\mathbf{x}_0)\label{newdipole}\end{eqnarray} 
where $\boldsymbol{\mu}_{m,n}=e\int d^3\boldsymbol{\xi}\psi_m^\ast(\boldsymbol{\xi})\boldsymbol{\xi}\psi_n(\boldsymbol{\xi})=\boldsymbol{\mu}_{n,m}^\ast$ denote some transition dipoles and $e<0$ is the electron charge. The model relies on the Openheimer approximation which separates the dynamical evolution into a center of mass motion $\mathbf{x}(t)$ (supposed here irrelevant since the atom is not moving) and a relative motion $\boldsymbol{\xi}(t)$ corresponding approximately to the electron motion with respect to the nuclei. Here we introduced
$\Delta(\mathbf{x}-\mathbf{x}_0)$ which is a narrow peaked distribution centered on the coordinate $\mathbf{x}_0$ associated with the center of mass of the atomic system and normalized as 
$\int d^3\mathbf{x}\Delta(\mathbf{x}-\mathbf{x}_0)=1$. $\Delta(\mathbf{x}-\mathbf{x}_0)$ reduces to the Dirac distribution $\delta^3(\mathbf{x}-\mathbf{x}_0)$ in the point-like dipole limit.  Still, in order to avoid singular divergences we will here keep $\Delta(\mathbf{x}-\mathbf{x}_0)$ finite.\\ 
\indent Inserting the total dipole density $\mathbf{P}=\mathbf{P}_\Psi+\mathbf{P}_{\textrm{diel.}}$ into the Lagrangian Eq.~\ref{1bb} leads to the Euler-Lagrange equations \cite{A,B} corresponding to Maxwell's equations in the medium with full polarization $\mathbf{P}$.
\begin{eqnarray}
\boldsymbol{\nabla}\times\mathbf{B}(\mathbf{x},t)=\frac{1}{c}\partial_t\mathbf{D}(\mathbf{x},t), &\boldsymbol{\nabla}\cdot\mathbf{D}(\mathbf{x},t)=0
\nonumber \\
\boldsymbol{\nabla}\times\mathbf{E}(\mathbf{x},t)=-\frac{1}{c}\partial_t\mathbf{B}(\mathbf{x},t), &\boldsymbol{\nabla}\cdot\mathbf{B}(\mathbf{x},t)=0\label{6}
\end{eqnarray} where $\mathbf{E}=\mathbf{D}-\mathbf{P}_\Psi-\mathbf{P}_{\textrm{diel.}}$.\\
\indent Similarly, we deduce using the Euler-Lagrange method a dynamical equation for the dielectric field  $\mathbf{X}_\omega(\mathbf{x},t)$: 
\begin{eqnarray}
\partial_t^2\mathbf{X}_\omega(\mathbf{x},t)+\omega^2\mathbf{X}_\omega(\mathbf{x},t)=\sqrt{\frac{2\sigma_\omega(\mathbf{x})}{\pi}}\mathbf{E}(\mathbf{x},t)\label{envi}\end{eqnarray} with here 
 $\mathbf{E}=\mathbf{D}-\mathbf{P}_\Psi-\mathbf{P}_{\textrm{diel.}}$.\\
\indent  For the Schrodinger fields we  must be more careful. Indeed by using the  Euler-Lagrange classical method we should deduce the evolution equations
\begin{eqnarray}
\dot{b}_m(t)=-i\frac{E_m}{\hbar}b_m(t)+\frac{i\bar{\textbf{E}}(t)}{\hbar}\cdot\sum_{n}\boldsymbol{\mu}_{n,m}^\ast b_n(t)\label{evol}
\end{eqnarray} with  $\bar{\textbf{E}}(t)=\int d^3\mathbf{E}(\mathbf{x},t)\mathbf{x}\Delta(\mathbf{x}-\mathbf{x}_0)\simeq \mathbf{E}(\mathbf{x}_0,t)$ the averaged field applied on the atomic dipole. A similar equation is obtained for the complex conjugate (Hermitian) field $b_m^\dagger(t)$. From this equation we deduce $\frac{d(\sum_{m}b_m^\dagger(t)b_m(t))}{dt}=0$ which means that the observable $N(t)=\sum_{m}b_m^\dagger(t)b_m(t)$ is a constant of motion. However, at that stage Eq.~\ref{evol} is not fully quantum since in the present formalism $\bar{\textbf{E}}(t)$ and $b_n(t)$ do not commute and the order of operators must be specified. To derive the equation of motion for the matter field we should therefore introduce the Hamiltonian of the system. Using the previous Lagrangian we derive the full Hamiltonian $H(t)$ reads: 
 \begin{eqnarray}
H(t)=\int d^3\mathbf{x}:\frac{\mathbf{B}(\mathbf{x},t)^2+\mathbf{E}(\mathbf{x},t)^2}{2}: +H_M(t)+H_{\Psi_a}(t)\nonumber\\  \label{totalHam}
\end{eqnarray} 
 where $H_M(t)=\int d^3\mathbf{x}\int_{0}^{+\infty}d\omega\hbar\omega\mathbf{f}^{\dagger}_\omega(\mathbf{x},t)\mathbf{f}_\omega(\mathbf{x},t)$ and $H_{\Psi_a}(t)=\sum_{m}E_mb_m^\dagger(t) b_m(t)$.
$H(t)$ is clearly an integral of motion as it can be for instance proven by using the Poynting theorem discussed in \cite{A}  and which reads:
$-\partial_t(\frac{\mathbf{B}^2+\mathbf{E}^2}{2})=\boldsymbol{\nabla}\cdot(c\mathbf{E}\times\mathbf{B})+\mathbf{J}\cdot\mathbf{E}$ where $\mathbf{J}=\partial_t\mathbf{P}$. Integrating the dissipated power $\int d^3\mathbf{x}\mathbf{J}\cdot\mathbf{E}$ leads, after some simple calculations which will not be repeated here, to $\frac{d}{dt}H(t)=-\oint_{\Sigma_\infty} d\boldsymbol{\Sigma}\cdot c \textbf{E}\times\textbf{B}\rightarrow 0$. This shows that the energy $E=H(t)$ is naturally conserved in the limit of an infinite integration volume and supposing that the field decays fast enough at infinity (an hypothesis which makes sense only with some additional physical assumptions). Moreover, to build a quantized  version  of this field theory we insert in Eq.~\ref{totalHam} the usual normal order product convention $:[...]:$ \cite{A,B} which removes some infinite spurious quantities in the energy (here the normal order means that we let  bosonic and fermionic operators to commute or respectively anticommute, additionally the fermionic and bosonic operators mutually commute).  Using the Heisenberg evolution law $i\hbar\frac{d}{dt}O(t)=[O(t),H(t)]$ for any quantum operator $O(t)$ we thus easily deduce once again Eqs.~\ref{6} and \ref{envi}. Furthermore, a rigorous application of commutation and anticommutation rules shows that Eq.~\ref{evol} is actually correct if the electric field operator is indeed positioned before the annihilation operator $b_n$ (taking into account this constraint it is also easy to prove the constancy of $N(t)$).\\
\indent We will now in the following consider only the case of the idealized two-level atom. For this we introduce the ground state $m=1$ and the excited state $m=2$ such as $E_2-E_1=\hbar\omega_{21}>0$. We have $N(t)=b_1^\dagger(t)b_1(t)+b_2^\dagger(t)b_2(t)=N(t_0)$ and we can therefore write  the Schrodinger part of the Hamiltonian $H_{\Psi_a}(t)$  as
  \begin{eqnarray}
	H_{\Psi_a}(t)=E_1 b_1^\dagger(t)b_1(t)+E_2 b_2^\dagger(t)b_2(t)\nonumber\\
	=\frac{E_1+E_2}{2}N(t)+ \frac{\hbar\omega_{21}}{2}\sigma_z(t)
\end{eqnarray}  with by definition $\sigma_z(t)=b_2^\dagger(t)b_2(t)-b_1^\dagger(t)b_1(t)$. We get for the evolution equation of $\sigma_z(t)$:
  \begin{eqnarray}
\dot{\sigma}_z(t)=\frac{-2i(\bar{\textbf{E}}(t)+\bar{\textbf{P}}_\Psi(t))\cdot\boldsymbol{\mu}_{1,2} }{\hbar}\sigma(t)\nonumber\\+\frac{2i(\bar{\textbf{E}}(t)+\bar{\textbf{P}}_\Psi(t))\cdot\boldsymbol{\mu}_{1,2}^\ast}{\hbar} \sigma^\dagger(t)\label{sigmaz}
\end{eqnarray} with by definition $\sigma(t)=b_1^\dagger(t)b_2(t)$ and $\sigma^\dagger(t)=b_2^\dagger(t)b_1(t)$. We emphasize that in this equation it is  the field $\bar{\textbf{E}}(t)+\bar{\textbf{P}}_\Psi(t)=\bar{\textbf{D}}(t)-\bar{\textbf{P}}_{\textrm{diel.}}(t)$ which plays a central role, with $\bar{\textbf{P}}_\Psi(t)=(\boldsymbol{\mu}_{1,2}\sigma(t)+\boldsymbol{\mu}_{1,2}^\ast\sigma^\dagger(t))\Delta(0)$ (since $\mathbf{P}_\Psi(\mathbf{x},t)=(\boldsymbol{\mu}_{1,2}\sigma(t)+\boldsymbol{\mu}_{1,2}^\ast\sigma^\dagger(t))\Delta(\mathbf{x}-\mathbf{x}_0)$). Importantly the field operator $\bar{\textbf{E}}(t)+\bar{\textbf{P}}_\Psi(t)$ commutes with atomic operators such  as $\sigma(t)$, $\sigma_z(t)$ and this makes the calculation easier.\\ 
\indent We similarly  introduce the evolution equation for $\sigma(t)$ as : 
 \begin{eqnarray}
\dot{\sigma}(t)=-i\omega_{21}\sigma(t)- \frac{i(\bar{\textbf{E}}(t)+\bar{\textbf{P}}_\Psi(t))\cdot\boldsymbol{\mu}_{1,2}^\ast }{\hbar}\sigma_z(t)\label{sigma}
\end{eqnarray}
which again involves the field $\bar{\textbf{E}}(t)+\bar{\textbf{P}}_\Psi(t)$.\\
%%%%%%%%%%%%%%%%%%%%%%%%%%
\section{Description of the full electric field operator in presence of a polarizable medium and a fluorescent dipole}
\indent At that stage  and before to solve the previous dynamical problem we should give a reminding concerning the formal structure of the fully quantized electric field operator  discussed in \cite{A,B}.\\
\indent We showed in \cite{B} that the 
total electric field evaluated at time $t$ and  at point $\mathbf{x}$ is given by  
\begin{eqnarray}
\mathbf{E}(\mathbf{x},t)=\mathbf{E}^{(v)}(\mathbf{x},t)\nonumber\\+\int_0^{t-t_0}d\tau\int d^3\mathbf{x'} \boldsymbol{\Delta}_v(\tau,\mathbf{x},\mathbf{x'}) \cdot\mathbf{P}(\mathbf{x'},t-\tau). \label{formal}
\end{eqnarray}
In this equation $t_0$ is an initial time (which can be sent to the remote past $t_0\rightarrow -\infty$ if needed),  $\mathbf{E}^{(v)}(\mathbf{x},t)=\mathbf{D}^{(v)}(\mathbf{x},t)$ denotes the field associated with free-space propagating photons existing in absence of any dielectric surrounding medium and atoms \cite{A,B}. The quantization of this free field expanded into plane wave modes (labeled  by the wave-vector $\mathbf{k}_\alpha$ and the transverse  polarization $\boldsymbol{\hat{\epsilon}}_{\alpha,j}$, with $j=1,2$) leads to the following operator expressions~\cite{B}:
 \begin{eqnarray}
\mathbf{E}^{(v)}(\mathbf{x},t)=\sum_{\alpha,j} -\sqrt{\frac{\hbar \omega_\alpha}{2}}c_{\alpha,j}^{(v)}(t)\hat{\mathbf{k}}_\alpha\times\boldsymbol{\hat{\epsilon}}_{\alpha,j}\Phi_\alpha(\mathbf{x})+ hcc.\nonumber\\ \label{121}
\end{eqnarray} with $\Phi_\alpha(\mathbf{x})=e^{i\mathbf{k}_\alpha\cdot\mathbf{x}}/\sqrt{V}$ ($V$ is  the infinite box volume of the Born von Karman modal expansion method) and where the modal expansion coefficients $c_{\alpha,j}^{(v)}(t)=c_{\alpha,j}^{(v)}(0)e^{-i\omega_\alpha t}$ 
(with $\omega_\alpha=c|\mathbf{k}_\alpha|$) satisfy the usual commutation relations for bosons (i.e. $[c_{\alpha,j}^{(v)}(t),c_{\alpha',j'}^{(v)\dagger}(t)]=\delta_{\alpha,\alpha'}\delta_{j,j'}$ etc...). Furthermore,  as shown in \cite{A,B} we have $c_{\alpha,j}^{(v)}(0)=c_{\alpha,j}(t_0)e^{-i\omega_\alpha t_0}$ where $t_0$ is the initial time mentioned before and  where $c_{\alpha,j}(t_0)$ is an operator associated with the total field acting at time $t_0$.  In the rest of this work we will write Eq.~\ref{121} in the more compact form 
\begin{eqnarray}\mathbf{E}^{(v)}(\mathbf{x},t)=\sum_{\alpha,j}\mathbf{E}_{\alpha,j}^{(v)}(\mathbf{x})c_{\alpha,j}^{(v)}(0)e^{-i\omega_\alpha t}+ hcc.\end{eqnarray}
which involves the transverse electric mode profiles  $\mathbf{E}_{\alpha,j}^{(v)}(\mathbf{x})$. We emphasize that the set of functions $\mathbf{E}_{\alpha,j}^{(v)}(\mathbf{x})$ constitute an orthogonal mode basis satisfying the condition $\int d^3 \mathbf{x}\mathbf{E}_{\alpha,j}^{(v)}(\mathbf{x})\cdot\mathbf{E}_{\alpha',j'}^{(v)\ast}(\mathbf{x})=\frac{\hbar \omega_\alpha}{2}\delta_{\alpha,\alpha'}\delta_{j,j'}$
\indent Moreover, the tensor $\boldsymbol{\Delta}_v(\tau,\mathbf{x},\mathbf{x'})$ seen in Eq.~\ref{formal} is a retarded dyadic Green propagator which was written $-\frac{\partial_\tau^2\mathbf{U}_v(\tau,\mathbf{x},\mathbf{x'}) }{c^2}$ in \cite{B}. Here it is explicitly obtained as an inverse Fourier integral \cite{B}: 
\begin{eqnarray}
\boldsymbol{\Delta}_v(\tau,\mathbf{x},\mathbf{x'})=\int_{-\infty}^{+\infty}\frac{d\omega}{2\pi} e^{-i\omega\tau} \frac{\omega^2}{c^2}\mathbf{G}_v(\mathbf{x},\mathbf{x'},\omega)
\end{eqnarray}
which vanishes for $\tau<0$ and involves the knowledge  of the  usual Green dyadic function~\cite{Girard1996,Novotny,B} $\mathbf{G}_v(\mathbf{x},\mathbf{x'},\omega)$ in vacuum, i.e.,  solution of the equation 
\begin{eqnarray}
\boldsymbol{\nabla}\times\boldsymbol{\nabla}\times\mathbf{G}_v(\mathbf{x},\mathbf{x'},\omega)-\frac{\omega^2}{c^2}\mathbf{G}_v(\mathbf{x},\mathbf{x'},\omega)
\nonumber\\=\mathbf{I}\delta(\mathbf{x}-\mathbf{x'}). \label{152}
\end{eqnarray}
Using the definition for the Fourier transform $\widetilde{\mathbf{E}}(\mathbf{x},\omega)=\int_{-\infty}^{+\infty}\frac{dt}{2\pi} \mathbf{E}(\mathbf{x},t)e^{+i\omega\tau} $ (and an equivalent formula for the dipole density $\widetilde{\mathbf{P}}(\mathbf{x},\omega)$) it is actually easier to write Eq.~\ref{formal} as
 \begin{eqnarray}
\widetilde{\mathbf{E}}(\mathbf{x},\omega)=\widetilde{\mathbf{E}}^{(v)}(\mathbf{x},\omega)+\int d^3 \mathbf{x'}\frac{\omega^2}{c^2}\mathbf{G}_v(\mathbf{x},\mathbf{x'},\omega\cdot\widetilde{\mathbf{P}}(\mathbf{x'},\omega), \nonumber\\ \label{formalbis}
\end{eqnarray}
where $\tilde{\mathbf{E}}^{(v)}(\mathbf{x},\omega)=\sum_{\alpha,j}[\mathbf{E}_{\alpha,j}^{(v)}(\mathbf{x})c_{\alpha,j}^{(v)}(0)\delta(\omega-\omega_\alpha)+\mathbf{E}_{\alpha,j}^{(v)\ast}(\mathbf{x})c_{\alpha,j}^{(v)\dagger}(0)\delta(\omega+\omega_\alpha)$.
 Now in the present problem involving a polarizable medium with complex permittivity $\widetilde{\varepsilon}(\mathbf{x},\omega)=\widetilde{\varepsilon}'(\mathbf{x},\omega)+i\widetilde{\varepsilon}''(\mathbf{x},\omega)$ we have 
 \begin{eqnarray}
\widetilde{\mathbf{P}}(\mathbf{x},\omega)=\widetilde{\mathbf{P}}_\Psi(\mathbf{x},\omega)+\widetilde{\mathbf{P}}^{(0)}(\mathbf{x},\omega)\nonumber\\
+(\widetilde{\varepsilon}(\mathbf{x},\omega)-1)\widetilde{\mathbf{E}}(\mathbf{x},\omega)
\label{24}
\end{eqnarray}  where $\widetilde{\mathbf{P}}^{(0)}(\mathbf{x},\omega)$ is the fluctuating dipole density introduced by Gruner and Welsch~\cite{Gruner1995,Gruner1996,Yeung1996} and given by~\cite{B}:
\begin{eqnarray}
\widetilde{\mathbf{P}}^{(0)}(\mathbf{x},\omega)=\sqrt{\frac{\hbar\tilde{\varepsilon}''(\mathbf{x},\omega)}{\pi}}\mathbf{f}_{\omega}^{(0)}(\mathbf{x},0)\Theta(\omega)\nonumber\\
+\sqrt{\frac{\hbar\tilde{\varepsilon}''(\mathbf{x},-\omega)}{\pi}}\mathbf{f}_{-\omega}^{(0)\dagger}(\mathbf{x},0)\Theta(-\omega).
\label{25}
\end{eqnarray} with $\mathbf{f}_{\omega}^{(0)}(\mathbf{x},0)$ a fluctuating dipolar term associated with the dielectric medium (we have $\mathbf{f}_{\omega}^{(0)}(\mathbf{x},t)=\mathbf{f}_{\omega}^{(0)}(\mathbf{x},0)e^{-i\omega t}$ and $\mathbf{f}_{\omega}^{(0)}(\mathbf{x},0)=\mathbf{f}_{\omega}^{(0)}(\mathbf{x},t_0)e^{i\omega t_0}$).  Therefore, it is specially convenient to introduce the total Green dyadic function~\cite{Girard1996,Novotny,B} $\mathbf{G}(\mathbf{x},\mathbf{x'},\omega)$ in the polarizable medium solution of the equation 
\begin{eqnarray}
\boldsymbol{\nabla}\times\boldsymbol{\nabla}\times\mathbf{G}(\mathbf{x},\mathbf{x'},\omega)-\frac{\omega^2}{c^2}\widetilde{\varepsilon}(\mathbf{x},\omega)\mathbf{G}(\mathbf{x},\mathbf{x'},\omega)
\nonumber\\=\mathbf{I}\delta(\mathbf{x}-\mathbf{x'}). \label{152biq}
\end{eqnarray} 
We thus rewrite Eq.~\ref{formalbis} as 
\begin{eqnarray}
\widetilde{\mathbf{E}}(\mathbf{x},\omega)=\widetilde{\mathbf{E}}^{(0)}(\mathbf{x},\omega)+\int d^3 \mathbf{x'}\frac{\omega^2}{c^2}\mathbf{G}(\mathbf{x},\mathbf{x'},\omega)\cdot\widetilde{\mathbf{P}}_{\textrm{eff.}}(\mathbf{x'},\omega), \nonumber\\ \label{formaltri}
\end{eqnarray} where $\widetilde{\mathbf{P}}_{\textrm{eff.}}(\mathbf{x},\omega)=\widetilde{\mathbf{P}}_\Psi(\mathbf{x},\omega)+\widetilde{\mathbf{P}}^{(0)}(\mathbf{x},\omega)$. Here the new electric field
operator $\widetilde{\mathbf{E}}^{(0)}(\mathbf{x},\omega)$ is a solution of  Maxwell's equations in the dielectric medium in absence of $ \widetilde{\mathbf{P}}_{\textrm{eff.}}(\mathbf{x'},\omega)$ and we have the integral relation \cite{B}
\begin{eqnarray}
\widetilde{\mathbf{E}}^{(0)}(\mathbf{x},\omega)=\widetilde{\mathbf{E}}^{(v)}(\mathbf{x},\omega)+\int d^3 \mathbf{x}\frac{\omega^2}{c^2}\mathbf{G}_v(\mathbf{x},\mathbf{x'},\omega)\nonumber\\ 
\cdot(\widetilde{\varepsilon}(\mathbf{x},\omega)-1)\widetilde{\mathbf{E}}^{(0)}(\mathbf{x}',\omega)\nonumber\\
=\widetilde{\mathbf{E}}^{(v)}(\mathbf{x},\omega)+\int d^3 \mathbf{x}\frac{\omega^2}{c^2}\mathbf{G}(\mathbf{x},\mathbf{x'},\omega)\nonumber\\ 
\cdot(\widetilde{\varepsilon}(\mathbf{x},\omega)-1)\widetilde{\mathbf{E}}^{(v)}(\mathbf{x}',\omega)  \label{formalquatre}
\end{eqnarray} where in the last equality we used the important Lippman-Schwinger integral relations
\begin{eqnarray}
\mathbf{G}(\mathbf{x},\mathbf{x'},\omega)=\mathbf{G}_v(\mathbf{x},\mathbf{x'},\omega)+\int d^3 \mathbf{x}\frac{\omega^2}{c^2}\mathbf{G}_v(\mathbf{x},\mathbf{x'},\omega)\nonumber\\ 
\cdot(\widetilde{\varepsilon}(\mathbf{x},\omega)-1)\mathbf{G}(\mathbf{x},\mathbf{x'},\omega)\nonumber\\
=\mathbf{G}_v(\mathbf{x},\mathbf{x'},\omega)+\int d^3 \mathbf{x}\frac{\omega^2}{c^2}\mathbf{G}(\mathbf{x},\mathbf{x'},\omega)
\nonumber\\ \cdot(\widetilde{\varepsilon}(\mathbf{x},\omega)-1)\mathbf{G}_v(\mathbf{x},\mathbf{x'},\omega). \nonumber\\  \label{formalcinq}
\end{eqnarray}
 Therefore, we see that the electric field $\widetilde{\mathbf{E}}^{(0)}(\mathbf{x},\omega)$ is completely determined by the knowledge of the  vacuum electric field   $\widetilde{\mathbf{E}}^{(v)}(\mathbf{x},\omega)$ through a linear equation. Clearly $\widetilde{\mathbf{E}}^{(0)}(\mathbf{x},\omega)$ can be also rewritten as 
\begin{eqnarray}
\widetilde{\mathbf{E}}^{(0)}(\mathbf{x},\omega)=\sum_{\alpha,j}[\mathbf{E}_{\alpha,j}^{(0)}(\mathbf{x})c_{\alpha,j}^{(v)}(0)\delta(\omega-\omega_\alpha)\nonumber\\+\mathbf{E}_{\alpha,j}^{(0)\ast}(\mathbf{x})c_{\alpha,j}^{(v)\dagger}(0)\delta(\omega+\omega_\alpha)\label{expansion}\end{eqnarray} where the functions $\mathbf{E}_{\alpha,j}^{(0)}(\mathbf{x})$ are the classical electric fields which are solutions of the scattering problem of a plane wave $\mathbf{E}_{\alpha,j}^{(v)}(\mathbf{x})$ with pulsation $\omega_\alpha$ by the polarizable medium. Therefore, we have
\begin{eqnarray}
\widetilde{\mathbf{E}}^{(0)}_{\alpha,j}(\mathbf{x})=\widetilde{\mathbf{E}}_{\alpha,j}^{(v)}(\mathbf{x})+\int d^3 \mathbf{x}\frac{\omega_\alpha^2}{c^2}\mathbf{G}_v(\mathbf{x},\mathbf{x'},\omega_\alpha)\nonumber\\ 
\cdot(\widetilde{\varepsilon}(\mathbf{x},\omega_\alpha)-1)\widetilde{\mathbf{E}}_{\alpha,j}^{(0)}(\mathbf{x}',\omega)\nonumber\\
=\widetilde{\mathbf{E}}_{\alpha,j}^{(v)}(\mathbf{x})+\int d^3 \mathbf{x}\frac{\omega_\alpha^2}{c^2}\mathbf{G}(\mathbf{x},\mathbf{x'},\omega_\alpha)\nonumber\\ 
\cdot(\widetilde{\varepsilon}(\mathbf{x},\omega_\alpha)-1)\widetilde{\mathbf{E}}_{\alpha,j}^{(v)}(\mathbf{x}').  \label{formalsept}
\end{eqnarray}We  point out that contrarily to $\widetilde{\mathbf{E}}^{(v)}_{\alpha,j}(\mathbf{x})$ the fields $\widetilde{\mathbf{E}}^{(0)}_{\alpha,j}(\mathbf{x})$ do not constitute in general  an orthogonal mode basis but this does not prevent us to use it for expanding the field $\widetilde{\mathbf{E}}^{(0)}(\mathbf{x},\omega)$ as in Eq.~\ref{expansion}. Before leaving this subsection it is important to give the field expression in the time domain which reads~\cite{B}:
\begin{eqnarray}
\mathbf{E}(\mathbf{x},t)=\mathbf{E}^{(0)}(\mathbf{x},t)\nonumber\\+\int_0^{ t-t_0}d\tau\int d^3\mathbf{x'} \boldsymbol{\Delta}(\tau,\mathbf{x},\mathbf{x'}) \cdot\mathbf{P}_{\textrm{eff.}}(\mathbf{x'},t-\tau)\label{machinchose}
\end{eqnarray}
 where like for the vacuum case the causal propagator $\boldsymbol{\Delta}(\tau,\mathbf{x},\mathbf{x'})$ is given by the inverse Fourier transform
\begin{eqnarray}
\boldsymbol{\Delta}(\tau,\mathbf{x},\mathbf{x'})=\int_{-\infty}^{+\infty}\frac{d\omega}{2\pi} e^{-i\omega\tau} \frac{\omega^2}{c^2}\mathbf{G}(\mathbf{x},\mathbf{x'},\omega),\end{eqnarray}
 which vanishes for $\tau<0$ \cite{B}.
%%%%%%%%%%%%%
\section{The Sipe approach and the Wigner-Weisskopf approximation}
\indent In order to solve the system of dynamical equations we will first use the Wigner-Weisskopf approximation~\cite{Wigner} as analyzed by J.~Sipe \cite{Sipe1995}. For this purpose we first consider the total Hamiltonian $H(t)$ written as
 \begin{eqnarray}
H(t)=\int d^3\mathbf{x}:\frac{\mathbf{B}(\mathbf{x},t)^2+\mathbf{D}(\mathbf{x},t)^2}{2}:+H_M(t)+H_{\Psi_a}(t)\nonumber\\-\int d^3\mathbf{x}\mathbf{D}(\mathbf{x},t)\cdot\mathbf{P}(\mathbf{x},t)\nonumber\\ 
\end{eqnarray} $H(t)$ can be formally separated into a non-interacting part $H^{(0)}=\int d^3\mathbf{x}:\frac{\mathbf{B}(\mathbf{x},t)^2+\mathbf{D}(\mathbf{x},t)^2}{2}:+H_M(t)+H_{\Psi_a}(t)$  and a coupling term $H^{(I)}(t)=-\int d^3\mathbf{x}\mathbf{D}(\mathbf{x},t)\cdot\mathbf{P}(\mathbf{x},t)$.  We now expand the quantum states $|\Psi(t)\rangle$ into  the mode basis associated with the  non interacting part $H^{(0)}(t)$. We will consider the problem of spontaneous emission of light by a two-level system and following Sipe~\cite{Sipe1995} we postulate that the quantum state at time $t_0=0$ reads
   \begin{eqnarray}
|I(0)\rangle=|0_1,1_2,0_m,0_P;t_0\rangle
\end{eqnarray}
where $0_1$ means that no electron is in the lower energy state $E_1$ and $1_2$ implies that there is an electron in the upper energy state $E_2$ (in the following the notation $|0_1,1_2,0_m,0_P;t_0\rangle$ and other similar ones mean that the vector is actually an eigenstate of a complete set of operator at time $t_0=0$). Since $N(t)=N(0)$ we see that  for $t\geq 0$ the system is in an eigenstate of $N(t)$ corresponding to the eigenvalue $N=+1$. Similarly $0_m$ and  $0_P$ mean that there is no polarization and photon excitation in the system at the initial time. 
 At time $t>t_0$ the atomic system evolves to its  ground state and we approximately have:  
 \begin{eqnarray}
|I(t)\rangle\simeq S(t)e^{-i\frac{E_1}{\hbar}t}|0_1,1_2,0_m,0_P;t_0\rangle\nonumber\\
+\int d^3\mathbf{x}\int_0^{+\infty}d\omega b_\omega(\mathbf{x},t) f_\omega(\mathbf{x},0)^\dagger|1_1,0_2,0_m,0_P;t_0\rangle\nonumber\\
+\sum_{\alpha,j}b_{\alpha,j}(t)c_{\alpha,j}(0)^\dagger|1_1,0_2,0_m,0_P;t_0\rangle+...\nonumber\\
\end{eqnarray} with $S(0)=1$ and $b_{\alpha,j}(0)=0$, $b_\omega(\mathbf{x},0)=0$. We also define a second possible state which we call the ground state of the complete system and that we write neglecting dressing and following Sipe~\cite{Sipe1995} as:
\begin{eqnarray}
|G(t)\rangle \simeq e^{-i\frac{E_1}{\hbar}t}|1_1,0_2,0_m,0_P;t_0\rangle\nonumber\\
=e^{-i\frac{E_1}{\hbar}t}|G(0)\rangle \label{fund}
\end{eqnarray} which characterizes a system with fundamental energy $E_1$ (i.e., neglecting dressing).\\
\indent  To solve Eq.~\ref{sigma} coupled to Maxwell's equations we here consider the matrix elements $\langle G(0)|A(t)|I(0)\rangle=\langle G(t)|A(0)|I(t)\rangle$ associated with the operator  $A(t)$ is expressed in the Heisenberg picture. We also remind that if $U(t,0)$ denotes the unitary evolution operator associated with the full Hamiltonian $H(t)$, we have $|I(t)\rangle=U(t,0)|I(0)\rangle$, $|G(t)\rangle=U(t,0)|G(0)\rangle$ and, therefore we have $A(t)=U^{-1}(t,0)A(0)U(t,0)$.  Using Eq.~\ref{sigma} we get 
\begin{eqnarray}
\frac{d}{dt}\langle G(0)|\sigma(t)|I(0)\rangle=-i\omega_{21}\langle G(0)|\sigma(t)|I(0)\rangle\nonumber\\
- \frac{i\boldsymbol{\mu}_{1,2}^\ast }{\hbar}\cdot\langle G(0)|\sigma_z(t)(\bar{\textbf{E}}(t)+\bar{\textbf{P}}_\Psi(t))|I(0)\rangle\label{sigmabis}
\end{eqnarray}
In order to solve this equation we should evaluate the different matrix elements involved.  First, we have \begin{eqnarray}
\langle G(0)|\sigma(t)|I(0)\rangle=\langle G(t)|\sigma(0)|I(t)\rangle\nonumber\\
\simeq e^{i\frac{E_1}{\hbar}t}\langle G(0)|\sigma(0)|I(t)\rangle\nonumber\\
=S(t)
\end{eqnarray}  where we have used Eq.~\ref{fund}. \\ \indent
Second, the matrix element 
\begin{eqnarray}
\langle G(0)|\sigma_z(t)(\bar{\textbf{E}}(t)+\bar{\textbf{P}}_\Psi(t))|I(0)\rangle\nonumber\\=\langle G(t)|\sigma_z(0)(\bar{\textbf{E}}(0)+\bar{\textbf{P}}_\Psi(0))|I(t)\rangle\nonumber\\
\simeq e^{i\frac{E_1}{\hbar}t}\langle G(0)|\sigma_z(0)(\bar{\textbf{E}}(0)+\bar{\textbf{P}}_\Psi(0))|I(t)\rangle\nonumber\\
\label{retrucu}
\end{eqnarray} can be rewritten as $-e^{i\frac{E_1}{\hbar}t}\langle G(0)|(\bar{\textbf{E}}(0)+\bar{\textbf{P}}_\Psi(0))|I(t)\rangle\simeq\langle G(0)|(\bar{\textbf{E}}(t)+\bar{\textbf{P}}_\Psi(t))|I(0)\rangle$
since the quantum state $|G(0)\rangle$ corresponds to the eigenvalue $\sigma_z=-1$.
Regrouping the terms we finally get the dynamical equation
\begin{eqnarray}
\frac{d}{dt}S(t)=-i\omega_{21}S(t)\nonumber\\
+ \frac{i\boldsymbol{\mu}_{1,2}^\ast }{\hbar}\cdot\langle G(0)|(\bar{\textbf{E}}(t)+\bar{\textbf{P}}_\Psi(t))|I(0)\rangle. \label{sigmatri}
\end{eqnarray} 
Now, to evaluate $\langle G(0)|\bar{\textbf{E}}(t)|I(0)\rangle$ we need  to insert the electric field solution of Maxwell's  equations evaluated at the dipole position. As discussed in the previous subsection the total field is given by Eq.~\ref{machinchose}.  By using the form for $\langle G(0)|$ and $|I(t)\rangle$ it is not difficult \cite{note} to show that we have $\langle G(0)|\bar{\textbf{E}}^{(0)}(t)|I(0)\rangle=0$, $\langle G(0)|\bar{\textbf{P}}^{(0)}(t)|I(0)\rangle=0$. Therefore, the only contribution to the electric field matrix element comes out from the field generated by the atomic dipole itself with:
\begin{eqnarray}
\langle G(0)|\bar{\textbf{P}}_\Psi(t)|I(0)\rangle=(\boldsymbol{\mu}_{1,2}\langle G(0)|\sigma(t)|I(0)\rangle\nonumber\\+\boldsymbol{\mu}_{1,2}^\ast\langle G(0)|\sigma^\dagger(t)|I(0)\rangle)\Delta(0)\nonumber\\
\simeq\boldsymbol{\mu}_{1,2}S(t)\Delta(0).\label{artics}
\end{eqnarray} where we have used  \begin{eqnarray}
\langle G(0)|\sigma^\dagger(t)|I(0)\rangle\simeq e^{i\frac{E_1}{\hbar}t}\langle G(0)|\sigma^\dagger(0)|I(t)\rangle=0.\nonumber \\
\end{eqnarray} 
 Regrouping all these expressions, Eq.~\ref{sigmatri} finally reads: 
\begin{eqnarray}
\frac{d}{dt}S(t)=-i\omega_0 S(t)
+ \int_0^{t}d\tau\frac{i\boldsymbol{\mu}_{1,2}^\ast\cdot\bar{\boldsymbol{\Delta}}(\tau,\mathbf{x}_0,\mathbf{x}_0)\cdot\boldsymbol{\mu}_{1,2}}{\hbar } \nonumber\\ \cdot S(t-\tau)\nonumber\\ \label{sigmafi}
\end{eqnarray} 
where $\omega_0=\omega_{21}-\frac{|\boldsymbol{\mu}_{1,2}|^2\Delta(0)}{\hbar}$ is a modified pulsation, due to the dipole $\langle G(0)|\bar{\textbf{P}}_\Psi(t)|I(0)\rangle$. This result is central for the present analysis since it shows that the vacuum photon field and the material fluctuating currents are not playing an effective role in the dynamical equation. Therefore, the transition dynamics is driven by the self-interaction of the source electromagnetic field. We emphasize that in Eq.~\ref{sigmafi} the Green dyadic tensor $\boldsymbol{\Delta}(\tau,\mathbf{x},\mathbf{x}_0)$ is actually a highly singular function both in the spatial  (near $\mathbf{x}_0$) and  in the time domain (near $\tau=0$). Actually, we showed in \cite{B}, using the Laplace transform method, that $\boldsymbol{\Delta}(\tau,\mathbf{x},\mathbf{x}_0)=[\textbf{Q}(\tau,\mathbf{x},\mathbf{x}_0)-\textbf{I}\delta^{3}(\mathbf{x}-\mathbf{x}_0)\delta(\tau)]\Theta(\tau)$ where $\textbf{Q}(\tau,\mathbf{x},\mathbf{x}_0)$ is a distribution that is regular in the time domain at $\tau=0$. Introducing this definition in   Eq.~\ref{sigmafi} shows that the integral $\int_0^{t}d\tau\textbf{Q}(\tau,\mathbf{x},\mathbf{x}_0)S(t-\tau)$ actually vanishes for $t\rightarrow 0^+$ and that the dipole correction
to $\omega_{21}$ seen in $\omega_0$ compensates exactly for the additional dipole term coming from the equality $\int_0^{t}d\tau\bar{\boldsymbol{\Delta}}(\tau,\mathbf{x}_0,\mathbf{x}_0)\cdot\boldsymbol{\mu}_{1,2}S(t-\tau)=\int_0^{t}d\tau\bar{\textbf{Q}}(\tau,\mathbf{x}_0,\mathbf{x}_0)\cdot\boldsymbol{\mu}_{1,2}S(t-\tau)-\langle G(0)|\bar{\textbf{P}}_\Psi(t)|I(0)\rangle$.\\
\indent Eq.~\ref{sigmafi}  can be solved more easily using the Laplace transform formalism applied to the field $S(t)$. Using Eq.~\ref{sigmafi} one obtains:  
\begin{eqnarray}
p\overline{S}(p)-S(0)=-i\omega_0\overline{S}(p)\nonumber\\
- \frac{i\boldsymbol{\mu}_{1,2}^\ast \cdot\frac{p^2}{c^2}\mathbf{G}(\mathbf{x}_0,\mathbf{x}_0,ip)\cdot\boldsymbol{\mu}_{1,2}}{\hbar}\overline{S}(p)\label{sigmafour}
\end{eqnarray} 
Therefore the following solution holds:
\begin{eqnarray}
S(t)=\int_{\gamma-i\infty}^{\gamma+i\infty}\frac{i dp}{2\pi}\frac{e^{pt}S(0)}{p+i\omega_0+\frac{i\boldsymbol{\mu}_{1,2}^\ast \cdot\frac{p^2}{c^2}\mathbf{G}(\mathbf{x}_0,\mathbf{x}_0,ip)\cdot\boldsymbol{\mu}_{1,2}}{\hbar}}.\nonumber\\ \label{rebukkk}
\end{eqnarray}
The calculation of this integral is given in the appendix A using the Wigner-Weisskopf approach.  In short, the idea is to assume for long time $t\gg 0$ the exponential decay law 
$S(t)=S(0)e^{-i\tilde{\omega}_0 t}$, where $\tilde{\omega}_0$ is a complex  frequency defined as $\tilde{\omega}_0=\omega_0-i\Gamma/2+\delta$, where $\Gamma\geq 0$ and $\delta$ are real numbers (we have also $S(0)=1$ by definition of the operator and quantum state). Now if we suppose that to a good approximation (called the polar approximation) we have 
\begin{eqnarray}
\tilde{\omega}_0=\omega_{0}-\frac{\boldsymbol{\mu}_{1,2}^\ast \cdot\frac{\omega_{0}^2}{c^2}\mathbf{G}(\mathbf{x}_0,\mathbf{x}_0,\omega_{0}+i0^+)\cdot\boldsymbol{\mu}_{1,2}}{\hbar}
\end{eqnarray}  which indeed justifies the decay law (see Appendix A). This allows us to write the decay rate
\begin{eqnarray}
\Gamma=2\textrm{Im}[\frac{\boldsymbol{\mu}_{1,2}^\ast \cdot\frac{\omega_0^2}{c^2}\mathbf{G}(\mathbf{x}_0,\mathbf{x}_0,\omega_{0}+i0^+)\cdot\boldsymbol{\mu}_{1,2}}{\hbar}]\label{gamma}
\end{eqnarray} and the lamb shift 
\begin{eqnarray}
-\delta=\textrm{Re}[\frac{\boldsymbol{\mu}_{1,2}^\ast \cdot\frac{\omega_0^2}{c^2}\mathbf{G}(\mathbf{x}_0,\mathbf{x}_0,\omega_{0}+i0^+)\cdot\boldsymbol{\mu}_{1,2}}{\hbar}]\nonumber\\
=\int_{-\infty}^{+\infty}\frac{d\omega}{\pi}P.V.[\frac{\textrm{Im}[\frac{\boldsymbol{\mu}_{1,2}^\ast \cdot\frac{\omega^2}{c^2}\mathbf{G}(\mathbf{x}_0,\mathbf{x}_0,\omega)\cdot\boldsymbol{\mu}_{1,2}}{\hbar}]}{\omega-\omega_{0}}]
\end{eqnarray} where we used  the Kramers-Kronig relation at the end. We point out that from the symmetry $\textrm{Im}[\mathbf{G}(\mathbf{x},\mathbf{x'},-\omega)]=-\textrm{Im}[\mathbf{G}(\mathbf{x},\mathbf{x'},\omega)]$ we have: 
\begin{eqnarray}
-\delta=\int_{0}^{+\infty}\frac{d\omega}{\pi}P.V.[\frac{\textrm{Im}[\frac{\boldsymbol{\mu}_{1,2}^\ast \cdot\frac{\omega^2}{c^2}\mathbf{G}(\mathbf{x}_0,\mathbf{x}_0,\omega)\cdot\boldsymbol{\mu}_{1,2}}{\hbar}]}{\omega-\omega_{0}}]\nonumber\\
+\int_{0}^{+\infty}\frac{d\omega}{\pi}\frac{\textrm{Im}[\frac{\boldsymbol{\mu}_{1,2}^\ast \cdot\frac{\omega^2}{c^2}\mathbf{G}(\mathbf{x}_0,\mathbf{x}_0,\omega)\cdot\boldsymbol{\mu}_{1,2}}{\hbar}]}{\omega+\omega_{0}}].   
\end{eqnarray} The first integral term is the correct Lamb shift obtained in the rotating wave approximation and the Wigner-Weisskopf theory. The second integral has actually the wrong sign  (see the analysis of the problem in \cite{Milonnibook}) and is clearly non resonant. Only a more precise theory going beyond the Wigner-Weisskopf polar approximation would justify this erroneous (small) value and this will not be considered here.\\
\indent  Moreover, the total field at point $\mathbf{x}$  would be in principle calculated  using: \begin{eqnarray}
\mathcal{E}(\mathbf{x},t):=\langle G(0)|\textbf{E}(\mathbf{x},t)|I(0)\rangle\nonumber\\=\int_{-\infty}^{+\infty}\frac{d\omega}{2\pi i}\frac{\frac{\omega^2}{c^2}\mathbf{G}(\mathbf{x},\mathbf{x_0},\omega)\cdot\boldsymbol{\mu}_{1,2}e^{-i\omega t}S(0)}{\omega_0-\omega-i0^+ -\frac{\omega_0^2}{\hbar c^2}\boldsymbol{\mu}_{1,2}^\ast\cdot\mathbf{G}(\mathbf{x},\mathbf{x_0},\omega)\cdot\boldsymbol{\mu}_{1,2}} \nonumber\\  \label{fieldencoreplusimportant}
\end{eqnarray}
However, the polar approximation allows us to define in a simpler form  this matrix element associated with  the spontaneously  emitted photon electric field:
  \begin{eqnarray}
\mathcal{E}(\mathbf{x},t):=\langle G(0)|\textbf{E}(\mathbf{x},t)|I(0)\rangle\nonumber\\=\int_{-\infty}^{+\infty}\frac{d\omega}{2\pi i}\frac{\frac{\omega^2}{c^2}\mathbf{G}(\mathbf{x},\mathbf{x_0},\omega)}{\tilde{\omega}_0-\omega-i0^+} \cdot\boldsymbol{\mu}_{1,2}e^{-i\omega t}S(0) \nonumber\\ \label{fieldimportant}
\end{eqnarray} 
The explicit calculation of  $\mathcal{E}(\mathbf{x},t)$ can only be done with the same conditions as used for evaluating $S(t)$ in Eq.~\ref{rebukkk}. For Eq.~\ref{rebukkk} it was necessary to suppose $t\gg t_0$. This is however not sufficient here and we will consider the far-field  far away from the source region where the propagator reads asymptotically : $\mathbf{G}(\mathbf{x},\mathbf{x_0},\omega_{0}+i0^+)\simeq \mathbf{F}(\mathbf{x},\mathbf{x_0},\omega_{0}+i0^+)e^{i\omega_{0}\sqrt{\tilde{\varepsilon}(\omega_{0})}R/c}$ where $R=|\mathbf{x}-\mathbf{x_0}|\gg c/\omega_{0}$ and where 
 $\mathbf{F}(\mathbf{x},\mathbf{x_0},\omega_{0}+i0^+)$ is a smoothly varying form factor characterizing the emission profile (for the permittivity we here suppose a background but this could be vacuum. Using this robust far-field approximation we have  
\begin{eqnarray}
\mathcal{E}(\mathbf{x},t)\simeq \frac{\omega_0^2}{c^2}\mathbf{F}(\mathbf{x},\mathbf{x_0},\omega_{0}+i0^+)  \cdot\boldsymbol{\mu}_{1,2}\nonumber\\ e^{-i\tilde{\omega}_0(t-\sqrt{\tilde{\varepsilon}(\omega_{0})}R/c)}S(0)\Theta(t-\sqrt{\tilde{\varepsilon}(\omega_{0})}R/c),\nonumber\\ \label{farfieldd}
\end{eqnarray} where the Heaviside function is reminiscent of the causal nature of the single photon emission (since the photon emission starts at $t_0=0$ no light exists outside the future-oriented light cone with apex located at $\mathbf{x_0}$, $t_0$) and is here justified by the nature of the Bromwich integral.\\
%%%%%%%%%%%
\section{Discussions}
\subsection{Local density of states and polaritonic wave functions}
\indent Some important remarks should be done here concerning the above derivation and its meaning.  First, as observed by Sipe,  $\mathcal{E}(\mathbf{x},t)$ is defining together with  $\mathcal{B}(\mathbf{x},t):=\langle G(0)|\textbf{B}(\mathbf{x},t)|I(0)\rangle$ a wave function for the single emitted photon. More precisely, starting from Maxwell's quantum equations for operators  $\textbf{E}(\mathbf{x},t)$ and $\textbf{B}(\mathbf{x},t)$  we can define some Maxwell's equations for the complex fields $\mathcal{E}(\mathbf{x},t)$ and $\mathcal{B}(\mathbf{x},t)$ which reads:
\begin{eqnarray}
\boldsymbol{\nabla}\times\mathcal{B}(\mathbf{x},t)=\frac{1}{c}\partial_t\mathcal{D}(\mathbf{x},t), &\boldsymbol{\nabla}\cdot\mathcal{D}(\mathbf{x},t)=0\nonumber\\
\boldsymbol{\nabla}\times\mathcal{E}(\mathbf{x},t)=-\frac{1}{c}\partial_t\mathcal{B}(\mathbf{x},t), &\boldsymbol{\nabla}\cdot\mathcal{B}(\mathbf{x},t)=0\label{6bb}
\end{eqnarray} In these equations  $\mathcal{D}(\mathbf{x},t)$ is defined as $\langle G(0)|\textbf{D}(\mathbf{x},t)|I(0)\rangle$ and involves  the complex polarization field  $\mathcal{P}(\mathbf{x},t):=\langle G(0)|\textbf{P}(\mathbf{x},t)|I(0)\rangle$. From \cite{A,B} and  Eq.~\ref{artics} we get 
 \begin{eqnarray}
\mathcal{P}(\mathbf{x},t)=\mathcal{P}_{\textrm{eff.}}(\mathbf{x},t)+\int_{0}^{t}\chi(\mathbf{x},\tau)d\tau\mathcal{E}(\mathbf{x},t-\tau),\nonumber\\
\label{18bisbis}
\end{eqnarray}  where $\chi(\mathbf{x},\tau)$ is the local linear susceptibility  of the inhomogeneous medium defined in \cite{A}. From Eq.~\ref{artics} we have
\begin{eqnarray}
\mathcal{P}_{\textrm{eff.}}(\mathbf{x},t)\simeq\boldsymbol{\mu}_{1,2}S(t)\Delta(\mathbf{x}-\mathbf{x}_0).\label{sourceterm}
\end{eqnarray} and $S(t)=e^{-i\tilde{\omega}_0 t}$ is the complex  valued dipole amplitude given in Eq.~\ref{rebukkk}. In other words, if we insert the source term given by  Eq.~\ref{sourceterm}  in the Maxwell equations Eq.~\ref{6bb} we can solve the problem directly using the propagator $\boldsymbol{\Delta}(\tau,\mathbf{x},\mathbf{x'})$ defined previously for the inhomogeneous dielectric problem. This solution is essentially classical and will automatically lead to Eq.~\ref{farfieldd} in the far-field of the quasi point-like dipole $\boldsymbol{\mu}_{1,2}S(t)$ associated with the  polarization density given by Eq.~\ref{sourceterm}.  The methods is associated with the first quantization approach of photon proposed by Sipe~\cite{Sipe1995} and Bialinicky-Birula~\cite{Birula} in which $\mathcal{D}$ and $\mathcal{B}$ define a wave function for the emitted photon.\\  
\indent A second, remark connected to the first one  deals with  the energy conservation and the meaning of $\Gamma$ in Eq.~\ref{gamma}. Indeed, the structure of this mathematical expression for $\Gamma$ is reminiscent of a classical calculation for the power radiated by an oscillating point-like dipole~\cite{Novotny}. This is clear since we can write $\Gamma$ as  
\begin{eqnarray}
\Gamma=\frac{\pi}{3}\frac{\omega_{0}}{\hbar}|\boldsymbol{\mu}_{1,2}|^2\rho_{LDOS}(\mathbf{x}_0)\label{gammab}
\end{eqnarray} 
  where 
	\begin{eqnarray}
\rho_{LDOS}(\mathbf{x}_0)=\frac{6\omega_{0}}{\pi c^2}\textrm{Im}[\hat{\textbf{n}}^\ast\cdot\mathbf{G}(\mathbf{x}_0,\mathbf{x}_0,\omega_{0}+i0^+)\cdot\hat{\textbf{n}}\label{LDOS} \nonumber\\
\end{eqnarray} with $\boldsymbol{\mu}_{1,2}=|\boldsymbol{\mu}_{1,2}|\hat{\textbf{n}}$. This is rigorously equivalent to the classical formula obtained for the power $P_0$ of a radiating dipole at the pulsation $\omega_0$ which reads 
\begin{eqnarray}
P_0=\frac{\pi}{3}\omega_{0}^2|\boldsymbol{\mu}_{1,2}|^2\rho_{LDOS}(\mathbf{x}_0).
\end{eqnarray} 
The last expression is identical to  Eq.~\ref{gammab} if we identify  the radiative power $P_0$ and the rate $\hbar\omega_0\Gamma$.  In order to give a justification to this identification we start from Eq.~\ref{6bb} and we obtain a complexified version of the Poynting theorem which reads:
 \begin{eqnarray}
-\partial_t(|\mathcal{B}|^2+|\mathcal{E}|^2)=2c\boldsymbol{\nabla}\cdot(\textrm{Re}[\mathcal{E}\times\mathcal{B}^\ast])+2\textrm{Re}[\mathcal{J}\cdot\mathcal{E}^\ast]\nonumber\\ \label{12bis},
\end{eqnarray} with $\mathcal{J}=\partial_t\mathcal{P}$ the complex dipolar current associated with Eq.~\ref{18bisbis}:
\begin{eqnarray}
\mathcal{J}(\mathbf{x},t)=\partial_t\mathcal{P}_{\textrm{eff.}}(\mathbf{x},t)+\int_{0}^{t}\chi(\mathbf{x},\tau)d\tau\partial_t\mathcal{E}(\mathbf{x},t-\tau)\nonumber\\+\chi(\mathbf{x},t)d\tau\mathcal{E}(\mathbf{x},0)\nonumber\\
\label{courantamer}
\end{eqnarray} 
 Moreover, by integration over the volume we can define the dissipated power inside the particle as $W_\Psi(t):=\int d^3\mathbf{x} 2\textrm{Re}[\partial_t\mathcal{P}_{\textrm{eff.}}^\ast(\mathbf{x},t)\cdot\mathcal{E}(\mathbf{x},t)]$. Using Eq.~\ref{sourceterm} we get:  
\begin{eqnarray}
W_\Psi(t)\simeq -2\textrm{Im}[\boldsymbol{\mu}_{1,2}^\ast \cdot\frac{\omega_{0}^3}{c^2}\mathbf{G}(\mathbf{x}_0,\mathbf{x}_0,\omega_{0}+i0^+)\nonumber\\ \cdot\boldsymbol{\mu}_{1,2}]e^{-\Gamma t}
=-\hbar\omega_{0}\Gamma e^{-\Gamma t}.
\end{eqnarray} This is with a minus sign the total radiated power $P_0$ discussed previously but weighted by the exponential decay factor $e^{-\Gamma t}$. Integrating $-W_\Psi(t)$ over time  we get the total energy emitted by the dipole from the initial time $t_0=0$ to time $t$: $\delta E=-\int_{0}^t dt' W_\Psi(t')=\hbar\omega_{0}(1-e^{-\Gamma t})$ which approaches $\hbar\omega_{0}$ if $t\rightarrow +\infty$. In evaluating  $W_\Psi(t)$ we used the fact that while $\mathbf{G}(\mathbf{x}_0,\mathbf{x}_0,\omega_{0}+i0^+)$ is a badly mathematically defined quantity this is not so for 
$\textrm{Im}[\mathbf{G}(\mathbf{x}_0,\mathbf{x}_0,\omega_{0}+i0^+)]$ which can be easily obtained by contour integration in the complex plane (see appendix B) and leads in the homogeneous surrounding medium case to:
 \begin{eqnarray}
\textrm{Im}[\mathbf{G}(\mathbf{x}_0,\mathbf{x}_0,\omega_{0}+i0^+)]=\frac{\omega_{0}}{6\pi c}\textrm{Re}[n_0(\omega_{0})]\textbf{I}.\label{regegege}
\end{eqnarray} where $n_0(\omega_{0})$ is the surrounding medium optical index. 
This allows us to define the LDOS in the general case and to justify directly Eq.~\ref{LDOS}. Furthermore, we have also
\begin{eqnarray}
\rho_{LDOS}(\mathbf{x}_0)=\frac{\omega_{0}^2}{\pi^2 c^3}\textrm{Re}[n_0(\omega_{0})]\nonumber\\
+\frac{6\omega_{0}}{\pi c^2}\textrm{Im}[\hat{\textbf{n}}^\ast\cdot\mathbf{G}_{\textrm{ref}}(\mathbf{x}_0,\mathbf{x}_0,\omega_{0}+i0^+)\cdot\hat{\textbf{n}}]
\end{eqnarray} where we have used a  standard separation\cite{Girard1996} of the Green tensor as $\mathbf{G}=\mathbf{G}_{\textrm{ref}}+\mathbf{G}_{0}$, where $\mathbf{G}_{0}$ is a contribution of the bulk medium of permittivity $\tilde{\varepsilon}(\omega_{0})=n_0^2(\omega_{0})$ and $\mathbf{G}_{\textrm{ref}}$ is an additional contribution originating from the inhomogeneities and various interfaces present in the system. \\
\indent  It is also important  for the present study to make a comment concerning the  theory of intensity measurement proposed by Glauber~\cite{Glauber,Milonnibook,Milonni2}. We remind that following the theory of Glauber  the photon detection rate  $I(\mathbf{x},t)$ at point $\mathbf{x}$ and time $t$ should generally be expressed as a convolution  between the temporal response of the detector $M(\tau)$ and the first-order correlation function of the electric field $\gamma(\mathbf{x},t,\tau)=\langle \textbf{E}^{(-)}(\mathbf{x},t)\textbf{E}^{(+)}(\mathbf{x},t-\tau)\rangle$, i.e.,    
\begin{eqnarray}
I(\mathbf{x},t)=2\textrm{Re}[\int_0^{+\infty}M(\tau) \gamma(\mathbf{x},t,\tau)].\label{rate}
\end{eqnarray} In the formula for $\gamma(\mathbf{x},t,\tau)$, $\textbf{E}^{(+)}(\mathbf{x},t)$  and $\textbf{E}^{(-)}(\mathbf{x},t)$ are respectively the positive and negative frequency operator  parts of the electric field containing  as usual only annihilation and creation operator for the photon field. In the broadband detector limit usually considered the formula simplifies and we get
 $I(\mathbf{x},t)\propto\langle \textbf{E}^{(-)}(\mathbf{x},t)\textbf{E}^{(+)}(\mathbf{x},t)\rangle$ which is the standard formula of Glauber~\cite{Glauber}. We stress that in order to derive Eq.~\ref{rate} a dipolar coupling  with  the detector was taken into account using a Hamiltonian interaction of the  usual form $H_{\textrm{int}}=-\textbf{p}\cdot \textbf{E}$ where $\textbf{p}$ is a dipole operator for the detector. However, from the point of view of the present dual formalism  the principal field to be  coupled to the detector is not the electric field $\textbf{E}$ but the displacement $\textbf{D}=\textbf{E}+\textbf{P}$ with $\textbf{P}$  the total dipole density of the medium (which in the interaction picture does not include the detector dipole contribution). As we showed in \cite{A,B} this displacement field is properly quantized by introducing a plane wave expansion with the general form (compare with Eq.~\ref{121}):   
\begin{eqnarray}
\mathbf{D}(\mathbf{x},t)=\sum_{\alpha,j} -\sqrt{\frac{\hbar \omega_\alpha}{2}}c_{\alpha,j}(t)\hat{\mathbf{k}}_\alpha\times\boldsymbol{\hat{\epsilon}}_{\alpha,j}\Phi_\alpha(\mathbf{x})+ hcc.\nonumber\\ \label{121new}
\end{eqnarray}
 where  $c_{\alpha,j}(t)$ and $c_{\alpha,j}^\dagger(t)$ are respectively the annihilation and creation operators associated with the photons in this dual formalism obeying usual commutation relations for bosons \cite{A,B}. Comparing with Eq.~\ref{121}  for  $\mathbf{E}^{(v)}(\mathbf{x},t)$
we see that the time dependency of  $c_{\alpha,j}(t)$ is not in general harmonic due to the coupling with the dipolar sources present~\cite{A,B}.  Using this description 
the positive frequency part $\mathbf{D}^{(+)}(\mathbf{x},t)$ of the displacement  field operator $\mathbf{D}(\mathbf{x},t)$ is clearly defined as 
\begin{eqnarray}
\mathbf{D}^{(+)}(\mathbf{x},t)=\sum_{\alpha,j} -\sqrt{\frac{\hbar \omega_\alpha}{2}}c_{\alpha,j}(t)\hat{\mathbf{k}}_\alpha\times\boldsymbol{\hat{\epsilon}}_{\alpha,j}\Phi_\alpha(\mathbf{x})\nonumber\\ \label{121newnew}
\end{eqnarray} and $\mathbf{D}^{(-)}(\mathbf{x},t)=\mathbf{D}^{(+)\dagger}(\mathbf{x},t)$ as usual.  In the dual formalism the interaction Hamiltonian  for the detection process actually reads $H^{\textrm{new}}_{\textrm{int}}=-\textbf{p}\cdot \textbf{D}$ and therefore the single photon rate correlation function is still given by Eq.~\ref{rate} with the correlation function now replaced by    
$\gamma(\mathbf{x},t,\tau)=\langle \textbf{D}^{(-)}(\mathbf{x},t)\textbf{D}^{(+)}(\mathbf{x},t-\tau)\rangle$. For all practical needs in the laboratory the use of $\textbf{D}$ instead of $\textbf{E}$ will not change anything since most single photon detectors are located in the far-field region, i.e., generally speaking in the air with $\textbf{D}\simeq \textbf{E}$. Still in the near-field regime the new formalism is in principle more powerful since it includes from the ground the lossy and dispersive dielectric environment.\\
\indent Furthermore, for the single photon process considered before we can write 
\begin{eqnarray}
\langle G(0)|\textbf{D}(\mathbf{x},t)|I(0)\rangle\simeq \langle G(0)|\textbf{D}^{(+)}(\mathbf{x},t)|I(0)\rangle
\end{eqnarray} where we used the fact that  the ground state is supposed here to be approximately the same at time $t=0$ and time  $t$.  Therefore, the recorded single photon intensity in the far-field by an idealized broadband detector requires only the knowledge of   $I(\mathbf{x},t)\propto\langle\textbf{D}^{(-)}(\mathbf{x},t)\textbf{D}^{(+)}(\mathbf{x},t)\rangle=|\langle G(0)|\textbf{D}^{(+)}(\mathbf{x},t)|I(0)\rangle|^2$ which from the previous analysis and Eq.~\ref{farfieldd} is given by  $|\mathcal{E}(\mathbf{x},t)|^2$. In agreement with Sipe's analysis in vacuum~\cite{Sipe1995} we thus finally obtain 
 a description of photon detection in terms of a single photon wave function in presence of a dielectric environment.\\
\subsection{The continuous regime and the optical Bloch equations}
\indent Before to conclude it is here important to study the continuous excitation regime when a laser mode interacts with the two-level atom considered previously. In order to find the optical Bloch equations in this regime we go back to Eqs.~\ref{sigmaz}, \ref{sigma} and study the dynamics of $\langle\sigma(t)\rangle$, and  $\langle\sigma_z(t)\rangle$ where the average is taken on an arbitrary initial state for the two-level atom. More specifically, we are interested in the evolution of $\langle\sigma(t)\rangle$, and  $\langle\sigma_z(t)\rangle$ with time under the influence of a quasi-classical electromagnetic wave characterized by an harmonic electric field $\mathbf{E}^{(0)}_L(\textbf{x},t)=\mathcal{E}^{0}_L(\textbf{x})e^{-\omega_Lt} +cc.$ ($\omega_L$ is the pulsation of the quasi-classical laser field). This electric field is a solution of the homogeneous classical Maxwell equations in presence of the dielectric medium. Therefore, from the point of view of the QED approach considered here it will  be necessary to include a contribution of the vacuum electric operator $\mathbf{E}^{(0)}(\textbf{x},t)$ in Eqs.~\ref{sigmaz}, \ref{sigma}. It is also clear that without the introduction of the operator $\mathbf{E}^{(0)}(\textbf{x},t)$ to preserve unitarity it would be also impossible to describe the excitation by an incident laser mode considered as a pure photonic state. With our formalism it is thus possible to describe the interaction process in complete analogy with what is done in the literature for an  atom excited by a laser beam in vacuum (i.e. without a dielectric lossy and dispersive surrounding).\\         
 \indent Now, lets start with Eq.~ \ref{sigma} and consider the average
	\begin{eqnarray}
\frac{d}{dt}\langle\sigma(t)\rangle=-i\omega_{21}\langle\sigma(t)\rangle- \frac{i\langle(\bar{\textbf{E}}(t)+\bar{\textbf{P}}_\Psi(t))\sigma_z(t)\rangle\cdot\boldsymbol{\mu}_{1,2}^\ast }{\hbar}\label{sigmaav}\nonumber\\
\end{eqnarray}
Here comes a difficulty because to solve this equation one must specify the operator ordering in $\langle(\bar{\textbf{E}}(t)+\bar{\textbf{P}}_\Psi(t))\sigma_z(t)\rangle$. This is a central issue which is well documented in the case of an atom in vacuum~\cite{Milonnibook}. The usual trick, that we should apply here as well, is to take a normal ordering in which  positive frequency part of the electric field are positioned to the right of  $\sigma_z$ while the negative frequency part of the electric field  operator  is positioned to the left of $\sigma_z$. This is allowed because atomic and field operators defined at the same time commute. If we can do that we will remove the contributions from $\mathbf{E}^{(0)}(\textbf{x},t)$  associated with vacuum  fluctuations and only study the effect of  the (classical) external interacting field and of the radiation-reaction.  We mention, that there are mathematical subtleties in 
the definition of positive and negative frequency parts and it exist actually two ways to define it which are not rigorously equivalent.  On the one side we could be tempted to consider a Fourier transform $\tilde{A}(\omega)$ of any operator $A(t)$ and thus define the positive frequency part  as $A^{(+)}(t)=\int_0^{+\infty} d\omega\tilde{A}(\omega)e^{-i\omega t}$ (similarly $A^{(-)}(t)=\int_{-\infty}^0 d\omega\tilde{A}(\omega)e^{-i\omega t}$). This way of defining $A^{(\pm)}(t)$ is actually correct if we have no interaction.  However, the canonical approach~\cite{Milonni2,Milonni3} is to use the separation between annihilation and creation operators for the fields and to define the positive frequency part by using only annihilation operators (respectively the negative frequency part is defined using only the creation operators for the fields). This is clearly the definition used for $\mathbf{D}^{(+)}(\mathbf{x},t)$ in Eq.~\ref{121newnew} but now this should be generalized for taking into account the relation $\bar{\textbf{E}}(t)+\bar{\textbf{P}}_\Psi(t)=\bar{\textbf{D}}(t)-\bar{\textbf{P}}_{\textrm{diel.}}(t)$. We give in Appendix C a detailed discussion of this important point in the present dual formalism.  We now write:            
\begin{eqnarray}
\langle(\bar{\textbf{E}}(t)+\bar{\textbf{P}}_\Psi(t))\sigma_z(t)\rangle\nonumber\\=\langle [(\bar{\textbf{E}}^{(+)}(t)+\bar{\textbf{P}}^{(+)}_\Psi(t))\sigma_z(t)
\nonumber\\+\langle\sigma_z(t)(\bar{\textbf{E}}^{(-)}(t)+\bar{\textbf{P}}^{(-)}_\Psi(t))]\rangle\nonumber\\
\end{eqnarray} with  for the dipole field operator $\bar{\textbf{P}}^{(+)}_\Psi(t)=\boldsymbol{\mu}_{1,2}\sigma(t)$ and  $\bar{\textbf{P}}^{(-)}_\Psi(t)=(\bar{\textbf{P}}^{(+)}_\Psi(t))^\dagger$. We have also for the electric field operators
\begin{eqnarray}
\mathbf{E}^{(\pm)}(\mathbf{x},t)=\mathbf{E}^{(0,\pm)}(\mathbf{x},t)\nonumber\\+\int_0^{ t-t_0}d\tau\int d^3\mathbf{x'} \boldsymbol{\Delta}^{(\pm)}(\tau,\mathbf{x},\mathbf{x'}) \cdot\mathbf{P}_{\textrm{eff.}}(\mathbf{x'},t-\tau)\label{machinchoseplus}.
\end{eqnarray} where the field $\mathbf{E}^{(0,\pm)}(\mathbf{x},t)=\mathcal{L}_t^{(\pm)}[\mathbf{E}^{(0,\pm)}(\mathbf{x},t)]$ and the dyadic $\boldsymbol{\Delta}^{(\pm)}(\tau,\mathbf{x},\mathbf{x'})=\mathcal{L}_\tau^{(\pm)}[\boldsymbol{\Delta}(\tau,\mathbf{x},\mathbf{x'})]$ are defined by applying the operator $\mathcal{L}_t^{(\pm)}=\frac{1}{2}[1\pm\frac{i\partial_t}{c\sqrt{-\boldsymbol{\nabla}^2}}]$ (see Appendix C and \cite{A}). As shown in Appendix D  a rigorous application of the operator $\mathcal{L}_t^{(\pm)}$ is in general difficult and an exact result is only obtained in  the vacuum. Fortunately, here we are interested in the dynamics at long time and we can use the approximation (see Eqs.~\ref{formalquatre}, and \ref{formalcinq}):
\begin{eqnarray}
\mathbf{E}^{(0,+)}(\mathbf{x},t)\simeq \int_0^{+\infty} d\omega\widetilde{\mathbf{E}}^{(0)}(\mathbf{x},\omega)e^{-i\omega t}\nonumber\\
\boldsymbol{\Delta}^{(+)}(\tau,\mathbf{x},\mathbf{x'})\simeq \int_0^{+\infty} \frac{d\omega}{2\pi}e^{-i\omega\tau} \frac{\omega^2}{c^2}\mathbf{G}(\mathbf{x},\mathbf{x'},\omega)\label{machinchoseplusplus}
\end{eqnarray} together with the relation $\mathbf{E}^{(0,-)}(\mathbf{x},t)=(\mathbf{E}^{(0,+)}(\mathbf{x},t))^\dagger$, $\boldsymbol{\Delta}^{(-)}(\tau,\mathbf{x},\mathbf{x'})=(\boldsymbol{\Delta}^{(+)}(\tau,\mathbf{x},\mathbf{x'}))^\ast$.\\
\indent  The next step is to remove the excitation from the initial state at time $t_0\rightarrow -\infty$. For this we use~\cite{Cohen2} for every operator the unitary transformation $A_{new}(t)=TA(t)T^{-1}$ where $T$ is the displacement operator defined as $T=\bigotimes_{\alpha,j} e^{[f^{(v)\ast}_{\alpha,j}(t)c_{\alpha,j}^{(v)}(t)-f_{\alpha,j}^{(v)}(t)c_{\alpha,j}^{(v)\dagger}(t)]}$ with $f_{\alpha,j}^{(v)}(t)=f_{\alpha,j}(t_0)e^{-i\omega_\alpha (t-t_0)}$ the modal coefficients in the expansion of the free laser field. More precisely, in analogy with Eq.~\ref{expansion} the laser field is written $\mathbf{E}^{(0)}_L(\mathbf{x},t)=\int_{-\infty}^{+\infty} d\omega\widetilde{\mathbf{E}}^{(0)}_L(\mathbf{x},\omega)e^{-i\omega t}$ with 
\begin{eqnarray}
\widetilde{\mathbf{E}}^{(0)}_L(\mathbf{x},\omega)=\sum_{\alpha,j}[\mathbf{E}_{\alpha,j}^{(0)}(\mathbf{x})f_{\alpha,j}^{(v)}(t_0)e^{i\omega_\alpha t_0}\delta(\omega-\omega_\alpha)+ cc.\nonumber\\
\label{expansionbis}\end{eqnarray}   
and where $\mathbf{E}_{\alpha,j}^{(0)}(\mathbf{x})$ is defined by  Eq.~\ref{formalsept} from the plane wave modes $\mathbf{E}_{\alpha,j}^{(v)}(\mathbf{x})$ (for the particular example used in this section  the laser is monochromatic so that we have necessarily $\omega_\alpha=\omega_L$). Moreover, within this formalism the application of $T$ on the initial coherent state $|L,t_0\rangle$ leads to the  photon vacuum: $T|L,t_0\rangle=|0,t_0\rangle$ and since $T$ acts only on the operator field $\mathbf{E}^{(0)}(\textbf{x},t)$ we deduce $T\mathbf{E}^{(0,+)}(\textbf{x},t)T^{-1}=\mathbf{E}^{(0,+)}(\textbf{x},t)+\mathbf{E}^{(0,+)}_L(\textbf{x},t)$, $T\mathbf{E}^{(0,-)}(\textbf{x},t)T^{-1}=\mathbf{E}^{(0,-)}(\textbf{x},t)+\mathbf{E}^{(0,-)}_L(\textbf{x},t)$. If we suppose that the initial quantum state is $|L,t_0\rangle\otimes|\textrm{atom},t_0\rangle$ (where $|\textrm{atom},t_0\rangle$ is any coherent superposition of the atomic states $|1,t_0\rangle$ and $|2,t_0\rangle$) then the coherent laser field can be removed from the quantum state (which now reads $|0,t_0\rangle\otimes|\textrm{atom},t_0\rangle$) and we should in turn add a classical laser field in the dynamical Eq.~\ref{sigmaav}. The previous analysis therefore generalizes the usual method for removing coherent states.  However, here the trick is now valid in presence of lossy and dispersive media.\\
\indent The rest of the derivation  is more conventional. We write $\sigma(t)=S(t)e^{-i\omega_L t}$ and using the rotating wave approximation we neglect contributions from  $\mathbf{E}^{(0,-)}_L$ and $\sigma^\dagger$. We get after some calculations:
 \begin{eqnarray}
\frac{d}{dt}\langle S(t)\rangle=-i(\omega_{0}-\omega_L)\langle S(t)\rangle \nonumber\\- \frac{i\bar{\mathcal{E}}^{(0)}_L\cdot\boldsymbol{\mu}_{1,2}^\ast }{\hbar}\langle\sigma_z(t)\rangle
+ N(t)
\label{sigmaav2}
\end{eqnarray}  
where  \begin{eqnarray}
N(t)= -i\int_0^{+\infty} \frac{d\omega}{2\pi}\frac{\omega^2}{c^2}\frac{\boldsymbol{\mu}_{1,2}^\ast\cdot\mathbf{G}(\mathbf{x}_0,\mathbf{x}_0,\omega)\cdot \boldsymbol{\mu}_{1,2}}{\hbar}\nonumber\\
\times \int_{-\infty}^{t-t_0}e^{i(\omega_L-\omega)\tau}\langle\sigma_z(t)S(t-\tau)\rangle \nonumber\\
\simeq +i\int_0^{+\infty} \frac{d\omega}{2\pi}\frac{\omega^2}{c^2}\frac{\boldsymbol{\mu}_{1,2}^\ast\cdot\mathbf{G}(\mathbf{x}_0,\mathbf{x}_0,\omega)\cdot \boldsymbol{\mu}_{1,2}}{\hbar}\nonumber\\
\times \int_{-\infty}^{t-t_0}e^{i(\omega_L-\omega)\tau}\langle S(t)\rangle .
\label{sigmaav2} 
\end{eqnarray} In the last equality we used the Markovian approximation~\cite{Milonnibook,Milonni3} $\langle\sigma_z(t)S(t-\tau)\rangle \simeq\langle\sigma_z(t)S(t)\rangle=-\langle S(t)\rangle $ (since $\sigma_z(t)\sigma(t)=-\sigma(t)$ by definition). In the long time limit with $t_0\rightarrow-\infty$ we have $N(t)=i\frac{\boldsymbol{\mu}_{1,2}^\ast\cdot\mathbf{G}(\mathbf{x}_0,\mathbf{x}_0,\omega)\cdot \boldsymbol{\mu}_{1,2}}{\hbar}\langle S(t)\rangle=(-\Gamma'/2-i\delta')$ where 
\begin{eqnarray}
\Gamma'=2\textrm{Im}[\frac{\boldsymbol{\mu}_{1,2}^\ast \cdot\frac{\omega_0^2}{c^2}\mathbf{G}(\mathbf{x}_0,\mathbf{x}_0,\omega_{L}+i0^+)\cdot\boldsymbol{\mu}_{1,2}}{\hbar}]\end{eqnarray} is the quantum rate at the laser frequency (compare with Eq.~\ref{gamma}) and \begin{eqnarray}
-\delta'=\int_{-\infty}^{+\infty}\frac{d\omega}{\pi}P.V.[\frac{\textrm{Im}[\frac{\boldsymbol{\mu}_{1,2}^\ast \cdot\frac{\omega^2}{c^2}\mathbf{G}(\mathbf{x}_0,\mathbf{x}_0,\omega)\cdot\boldsymbol{\mu}_{1,2}}{\hbar}]}{\omega-\omega_{L}}]
\end{eqnarray} is the new Lamb shift. We emphasize that for  most applications the difference between $\Gamma$ and $\Gamma'$, $\delta$ and $\delta'$ can be neglected. Therefore, within the general approach considered we obtained the first Bloch equation
 \begin{eqnarray}
\frac{d}{dt}\langle S(t)\rangle=-i(\omega_{0}-\delta-\omega_L)\langle S(t)\rangle \nonumber\\- \frac{i\bar{\mathcal{E}}^{(0)}_L\cdot\boldsymbol{\mu}_{1,2}^\ast }{\hbar}\langle\sigma_z(t)\rangle
-\Gamma/2\langle S(t)\rangle
\label{sigmaav3}
\end{eqnarray} which is now valid in presence of lossy and dispersive linear media. It is possible using the same procedure to deduce the second optical Bloch equation which reads within tthe same approximations:
 \begin{eqnarray}
\frac{d}{dt}\langle \sigma_z(t)\rangle=\frac{-2i\bar{\mathcal{E}}^{(0)\ast}_L\cdot\boldsymbol{\mu}_{1,2} }{\hbar}\langle S(t)\rangle\nonumber\\+\frac{2i\bar{\mathcal{E}}^{(0)}_L\cdot\boldsymbol{\mu}_{1,2}^\ast}{\hbar} \langle S^\dagger(t)\rangle \label{sigmazbis}
-\Gamma(1+\langle \sigma_z(t)\rangle).\nonumber\\
\end{eqnarray}
Therefore, we can by using the generalized Huttner Barnett model justify the use of optical Bloch equations which were often introduced with the more phenomenological Langevin noise approach
~\cite{Dzotjan2010,Cano2011,Hummer2013,Chen2013,Delga2014,Hakami2014,Choquette2012,Grimsmo2013,Rousseaux2016}. 
\section{Final remarks and conclusion}
\indent To conclude, we provided a description of spontaneous emission for a fluorescent two level atom using the generalized Hutner-Barnett approach given in \cite{A,B}. We showed that within this Hamiltonian description it is clearly possible to analyze rigorously spontaneous emission in a lossy and dispersive  inhomogeneous dielectric environment. Importantly, we showed that  while our description used the complete electromagnetic field including photon vacuum fluctuations and Langevin's noise current associated with the dielectric environment the spontaneous emission process  can be understood as resulting from a self-coupling of the  fluorescent dipole. In this perspective the LDOS appears thus as a consequence of the classical radiation reaction (in agreement with semi-classical approaches neglecting the quantization of losses~\cite{Colas2,Novotny}). On the one side, spontaneous emission is thus interpreted rather classically as a radiation reaction due to the Lorentz force~\cite{Milonnibook}. However, on the other side the full unitarity of quantum mechanics is respected in our formalism in order to preserve the canonical equal-time commutation relations. Therefore, as already discussed by Milonni in the context of photon/atoms coupling  in vacuum~\cite{Milonnibook} the description is not univocal and depends on the order we introduce operators in the dynamical equations. Here, the choice was done in order to favor the classical radiation force interpretation  but other choices are clearly possible and all of them are  equivalent.  Furthermore, this fact can be seen as a direct  consequence of the preservation of unitarity in our description. Without the inclusion of  both photon vacuum and fluctuating currents the alternative representations would not exist and the full unitarity would be broken. The present work justifies semi-classical results \cite{Colas2,Novotny,Drezet} and alternative quantum approaches based on the Langevin's noise method~\cite{Dzotjan2010,Cano2011,Hummer2013,Chen2013,Delga2014,Hakami2014,Choquette2012,Grimsmo2013,Rousseaux2016} which neglected the role of photon vacuum. We think that the present work will motivate further studies in order to analyze other regimes of coupling between emitters and dielectric media and will impact our description of quantum polaritonic an plasmonic physics in the quantum regime (e.g., with near-field optical microscopes involving single-photon emitters~\cite{Cuche2010,OrianePRB,Pham,Martin,Cuche2017}.             
%%%%%%%%%%%%%%%       
\section{Acknowledgments}
\indent This work was supported by Agence Nationale de la Recherche (ANR), France, through the PLACORE (ANR-13-BS10-0007) grant. The author gratefully acknowledges discussions with S. Huant.  
%%%%%%%%%%%%%%%%%%%%%%%%%%%%%%%%%%%%%%%%%%%%%%%%%%%%%%%%%%%%%%%%%%%%%%%%%%%%%%%%%%%%%%%%%%%%%%%appendix%%%%%%%%%%%%%%%%%%%%
%%%%%%%%%%%%%%%%%%%%%%%%%%%%%%%%%%%%%%%%%%%%%%%%%%%%%%%%%%%%%%%%%%%%%%%%%%%%%%%%%%%%%%%%%%%%%%%%%%%%%%%%%%%%%%%%%%%%%%%%%%%%%
\appendix
\section{The Wigner-Weisskopf approximation}
The evaluation of the integral Eq.~\ref{rebukkk} is in general difficult and we will here use the method proposed by Wigner and Weisskopf \cite{Wigner}. For this we introduce the notation $\overline{\beta}(p)=-\frac{\boldsymbol{\mu}_{1,2}^\ast \cdot\frac{p^2}{c^2}\mathbf{G}(\mathbf{x},\mathbf{x'},ip)\cdot\boldsymbol{\mu}_{1,2}}{\hbar}$ and we rewrite Eq.~\ref{sigmafour} as $\dot{S}(t)=-i\omega_0S(t)+i\int_0^{t}d\tau \beta(\tau)S(t-\tau)=-i\omega_{21}S(t)+i\int_0^{t}d\tau q(\tau)S(t-\tau)$ where $q(\tau)=\beta(\tau)+(\omega_{21}-\omega_0)\delta(\tau)$ is regular at time $\tau=0$. The ansatz made by Wigner and Weisskopf is to suppose for time $t\rightarrow +\infty$ an exponential decay $S(t)=S(0)e^{-i\tilde{\omega}_0 t}$ where $\tilde{\omega}_0$ is a complex  frequency which is defined as $\tilde{\omega}_0=\omega_0-i\Gamma/2+\delta$ with $\Gamma\geq 0$ and $\delta$ real numbers. This leads to the relation 
\begin{eqnarray}
\int_0^{t}d\tau \beta(\tau)e^{i\tilde{\omega}_0\tau}=i\Gamma/2-\delta\nonumber\\
=e^{i\tilde{\omega}_0 t}\int_{\gamma-i\infty}^{\gamma+i\infty}\frac{i dp}{2\pi}\frac{e^{p t}\overline{\beta}(p)}{p+i\tilde{\omega}_0}
\end{eqnarray}
or equivalently using $p=\gamma-i\omega$ and defining $B(\omega+i\gamma):=\overline{\beta}(p)$:
\begin{eqnarray}
i\Gamma/2-\delta
=e^{i\tilde{\omega}_0 t}\int_{-\infty}^{+\infty}\frac{d\omega}{2\pi}\frac{e^{-i\omega t)}B(\omega)}{i(\tilde{\omega}_0-\omega)}\nonumber\\
\end{eqnarray} in the limit $\gamma\rightarrow 0^+$. This integral is calculated by contour integration in the complex plane and taking into account that the poles of $B(\omega)$ are all located in the lower frequency half-plane (the derivation is identical to the one for $\tilde{\varepsilon}(\omega)$ as shown  in \cite{B}). Using the residue theorem we get
 \begin{eqnarray}
i\Gamma/2-\delta
=B(\tilde{\omega}_0)-\sum_{m}\frac{\textrm{res}[B(\Omega_m)]e^{i(\tilde{\omega}_0-\Omega_m) t}}{\tilde{\omega}_0-\Omega_m}\nonumber\\
\end{eqnarray}  where the sum is taken over the residues or poles $\Omega_m$ of $B(\omega)$. This equality cannot be valid at every time since the left hand side is independent of $t$ while the right hand side depends explicitly on $t$. Actually, the exponential decay law is only valid for long time, i.e., $t\gg 0$. However, the equality between the right hand side and  $\int_0^{t}d\tau \beta(\tau)e^{i\tilde{\omega}_0\tau}$ is valid at every time $t\geq0$ and since $\int_0^{t}d\tau q(\tau)e^{i\tilde{\omega}_0\tau}$  vanishes for $t=0$ we deduce 
    \begin{eqnarray}
B(\tilde{\omega}_0)=\sum_{m}\frac{\textrm{res}[B(\Omega_m)]}{\tilde{\omega}_0-\Omega_m}+\omega_0-\omega_{21}\nonumber\\ \label{polaire}
\end{eqnarray} which allows us to rewrite 
\begin{eqnarray}
i\Gamma/2-\delta
=\lim_{t\to\infty}\sum_{m}\textrm{res}[B(\Omega_m)]\left(\frac{1-e^{i(\tilde{\omega}_0-\Omega_m) t}}{\tilde{\omega}_0-\Omega_m}\right)\nonumber\\ +\omega_0-\omega_{21}\label{redur}
\end{eqnarray}
To conclude this evaluation we observe that in the limit where there is a continuum of poles $\Omega_m$  we can with a good approximation~\cite{Holstein}
write \begin{eqnarray}
\sum_{m}\textrm{res}[B(\Omega_m)]\frac{e^{i(\tilde{\omega}_0-\Omega_m) t}}{\tilde{\omega}_0-\Omega_m}\nonumber\\ \simeq 2\pi i e^{-i\tilde{\omega}_0 t}\sum_{m}\textrm{res}[B(\Omega_m)]\delta(\omega_0-\Omega'_m)\label{dur}
\end{eqnarray}
Furthermore, in the limit where the poles are near the real axis and where $\Gamma/2\gg -\Omega''_m$ we have:
 \begin{eqnarray}
B(\tilde{\omega}_0)\simeq B(\omega_0-i0^+)\simeq \sum_{m}\textrm{res}[B(\Omega_m)]P.V.[\frac{1}{\omega_0-\Omega'_m}]\nonumber\\+i\pi\sum_{m}\textrm{res}[B(\Omega_m)]\delta(\omega_0-\Omega'_m) +\omega_0-\omega_{21}\nonumber\\\label{tropdur}
\end{eqnarray} 
Inserting Eqs.~\ref{dur} and \ref{tropdur} in Eq.~\ref{redur} leads to: 
\begin{eqnarray}
i\Gamma/2-\delta\simeq\sum_{m}\textrm{res}[B(\Omega_m)]P.V.[\frac{1}{\omega_0-\Omega'_m}]\nonumber\\-i\pi\sum_{m}\textrm{res}[B(\Omega_m)]\delta(\omega_0-\Omega'_m) +\omega_0-\omega_{21}\label{moli}
\end{eqnarray} 
However from the definition of $B(z)$, with $z$ a complex number, we have $B(-z^\ast)=B^\ast (z)$. Therefore, for $\omega_0\pm i0^+$ we have after separating the real from the imaginary part $B'(\omega_0+i0^+)=B'(\omega_0-i0^+)$ and $B''(\omega_0+i0^+)=-B''(\omega_0-i0^+)$. This implies
 \begin{eqnarray}
B(\omega_0+i0^+)\simeq\sum_{m}\textrm{res}[B(\Omega_m)]P.V.[\frac{1}{\omega_0-\Omega'_m}]\nonumber\\-i\pi\sum_{m}\textrm{res}[B(\Omega_m)]\delta(\omega_0-\Omega'_m) +\omega_0-\omega_{21} \label{tropdurencore}
\end{eqnarray} (not the difference of sign with Eq.~\ref{tropdur}) and we get after comparison wih Eq.~\ref{moli}
\begin{eqnarray}
i\Gamma/2-\delta\simeq B(\omega_0+i0^+)\nonumber\\
=\frac{\boldsymbol{\mu}_{1,2}^\ast \cdot\frac{\omega_0^2}{c^2}\mathbf{G}(\mathbf{x},\mathbf{x'},\omega_0+i0^+)\cdot\boldsymbol{\mu}_{1,2}}{\hbar}
\end{eqnarray} which is the  final result.
%%%%%%%%%%%%%%%%%%%%%%%%%%%%%%
\section{The radiated power and the LDOS of a quantum dipole}
In order to calculate $W_\Psi(t)$ we have to consider the local field in the vicinity of the dipole. Using Eq.~\ref{fieldimportant}  we get for $\mathbf{x}\rightarrow\mathbf{x_0}$:
\begin{eqnarray}
\mathcal{E}(\mathbf{x},t)\simeq \frac{\omega_0^2}{c^2}\mathbf{G}(\mathbf{x},\mathbf{x_0},\omega_0+i0^+)\cdot\boldsymbol{\mu}_{1,2}e^{-i\tilde{\omega}_0 t}\nonumber\\ \label{gglgogo}
\end{eqnarray}
Moreover, the normed and finite function $\Delta(\mathbf{x}-\mathbf{x_0})$ characterizing the dipole polarization prevents us to obtain divergence in evaluating $W_\Psi(t)$. In the case of an homogeneous medium the Green tensor $\mathbf{G}(\mathbf{x},\mathbf{x_0},\omega_0+i0^+)$ is easily obtained as an integral over polarization states and wavevectors. We get (see \cite{Cohen}):
\begin{eqnarray}
\mathbf{G}(\mathbf{x}_0,\mathbf{x}_0,\omega_0+i0^+)\nonumber\\
=\int \frac{d^3\mathbf{k}}{(2\pi)^3}\frac{1}{k^2-\frac{\omega_0^2\tilde{\varepsilon}(\omega_0)}{c^2}}(\textbf{I}-\frac{\textbf{k}\otimes\textbf{k}}{\frac{\omega_0^2\tilde{\varepsilon}(\omega_0)}{c^2}})\nonumber\\
=\int_{-\infty}^{+\infty} \frac{kdk}{(2\pi)^3}\frac{2\pi \textbf{I} f_{\textrm{reg}}(k)}{k-\frac{\omega_0\sqrt{\tilde{\varepsilon}(\omega_0)}}{c}}(1-\frac{2}{3}\frac{k^2}{\frac{\omega_0^2\tilde{\varepsilon}(\omega_0)}{c^2}})\nonumber\\
\label{104bbbbis}
\end{eqnarray} In the last line we introduced a regularization function $f_{\textrm{reg}}(k)$ such as $f_{\textrm{reg}}(k)\simeq 1$ for value  near $k\simeq 0$ but  $f_{\textrm{reg}}(k)\simeq 0$ si $|k|\rightarrow +\infty$. This trick prevents the divergence and allows us to calculate the integral along a  contour in the upper part of the complex plane. Using the residue theorem one get finally 
   \begin{eqnarray}
\mathbf{G}(\mathbf{x}_0,\mathbf{x}_0,\omega_0+i0^+)\nonumber\\=i\frac{\omega_0}{6\pi c}\sqrt{\tilde{\varepsilon}(\omega_0)}\textbf{I}f_{\textrm{reg}}(\frac{\omega_0\sqrt{\tilde{\varepsilon}(\omega_0)}}{c}) .
\label{104bbbb}
\end{eqnarray}
At the end we can simplify since $f_{\textrm{reg}}(\frac{\omega_0\sqrt{\tilde{\varepsilon}(\omega_0)}}{c})\simeq 1$. We point out that this result  depends on the assumption concerning the convergence of 
$f_{\textrm{reg}}(k)$ at infinity. It is well known that the Green dyadic propagator is badly defined at the spatial origin and this is clearly another manifestation of this fact. Still the result concerning the imaginary part of $\mathbf{G}(\mathbf{x}_0,\mathbf{x}_0,\omega_0+i0^+)$ is very robust and will keep its absolute meaning since  only the real part contains potential divergences. Finally, taking the imaginary part of Eq.~\ref{104bbbb}   leads directly to the result Eq.~\ref{regegege}. 
%%%%%%%%%
\section{Definition of the positive and negative frequency parts of the electric field operator}
As we explained in  the Appendix B  of \cite{A} (see also \cite{Marx}) the positive (respectively negative) frequency part of the electric displacement operator $\mathbf{D}^{(\pm)}(\mathbf{x},t)$ is given by: 
\begin{eqnarray}
\mathbf{D}^{(\pm)}(\mathbf{x},t)=\mathcal{L}_t^{(\pm)}[\mathbf{D}(\mathbf{x},t)]
\end{eqnarray}
where  we have defined~\cite{Marx} the operator $\mathcal{L}_t^{(\pm)}=\frac{1}{2}[1\pm\frac{i\partial_t}{c\sqrt{-\boldsymbol{\nabla}^2}}]$. With this definition we get Eq.~\ref{121new} which ensures a separation between annihilation and creation operators, i.e., between  terms containing $c_{\alpha,j}(t)$ and those containing $c^\dagger_{\alpha,j}(t)$. 
Similarly, for the electric dipole density $\bar{\textbf{P}}_{\textrm{diel.}}(t)$ we introduce a separation between positive and negative frequency part by using the definition
$\mathbf{P}^{(+)}_{\textrm{diel.}}(\mathbf{x},t)=\mathcal{M}_t^{(\pm)}[\mathbf{P}_{\textrm{diel.}}(\mathbf{x},t)]$ where the linear operator $\mathcal{M}_t^{(\pm)}$ act on the material oscillator fields  $\mathbf{X}_\omega(\mathbf{x},t)$ in order to separate the contribution containing  annihilation operators $\mathbf{f}_\omega(\mathbf{x},t)$ from  the contribution containing only creation  operators $\mathbf{f}^\dagger_\omega(\mathbf{x},t)$. Moreover, as shown in \cite{A,B}  we have $\mathbf{f}_\omega(\mathbf{x},t)=\frac{i\partial_t\mathbf{X}_\omega(\mathbf{x},t)+\omega\mathbf{X}_\omega(\mathbf{x},t)}{\sqrt{2\hbar\omega}}$ and 
 \begin{eqnarray}
\mathbf{P}_{\textrm{diel.}}(\mathbf{x},t)=\int_0^{+\infty}d\omega\sqrt{\frac{\hbar\sigma_{\omega}(\mathbf{x})}{\pi\omega}}[\mathbf{f}_{\omega}(\mathbf{x},t)
+\mathbf{f}^{\dagger}_{\omega}(\mathbf{x},t)]. \nonumber\\ \label{didi}
\end{eqnarray}
Therefore if we define
\begin{eqnarray}
\mathbf{P}_{\textrm{diel.}}^{(+)}(\mathbf{x},t)=\int_0^{+\infty}d\omega'\int d^{3}\mathbf{x}'\mathbf{f}_{\omega'}(\mathbf{x}',t)\nonumber\\
\cdot[\mathbf{f}_{\omega'}(\mathbf{x}',t),\mathbf{P}_{\textrm{diel.}}(\mathbf{x},t)]
\end{eqnarray}
and use the canonical commutation relations \cite{A} we get:
\begin{eqnarray}
\mathbf{P}_{\textrm{diel.}}^{(+)}(\mathbf{x},t)
=\int_0^{+\infty}d\omega\sqrt{\frac{\hbar\sigma_{\omega}(\mathbf{x})}{\pi\omega}}\mathbf{f}_{\omega}(\mathbf{x},t)
\nonumber\\ \label{didibis}
\end{eqnarray} which defines the positive frequency part $\mathbf{P}_{\textrm{diel.}}^{(+)}(\mathbf{x},t)$ and the operator $\mathcal{M}_t^{(\pm)}$ (for the negative frequency part we simply use $\mathbf{P}_{\textrm{diel.}}^{(-)}(\mathbf{x},t)=(\mathbf{P}_{\textrm{diel.}}^{(+)}(\mathbf{x},t))^\dagger$).\\ 
\indent Finally, if we use the definition $\mathbf{P}_\Psi(\mathbf{x},t)=(\boldsymbol{\mu}_{1,2}\sigma(t)+\boldsymbol{\mu}_{1,2}^\ast\sigma^\dagger(t))\Delta(\mathbf{x}-\mathbf{x}_0)$ we can obtain a separation between the operators $\sigma(t)$ and $\sigma^\dagger(t)$ and we thus define $\mathbf{P}^{(\pm)}_\Psi(\mathbf{x},t)=\mathcal{N}_t^{(\pm)}[\mathbf{P}_\Psi(\mathbf{x},t)]$ by 
\begin{eqnarray}
\mathbf{P}^{(+)}_\Psi(\mathbf{x},t)=\boldsymbol{\mu}_{1,2}\sigma(t)\Delta(\mathbf{x}-\mathbf{x}_0)
\end{eqnarray} and $\mathbf{P}^{(-)}_\Psi=\mathbf{P}^{(+)}_\Psi()^\dagger$.\\
\indent Now, for the electric field we have  by definition $\mathbf{E}(\mathbf{x},t)=\mathbf{D}(\mathbf{x},t)-\mathbf{P}_{\textrm{diel.}}(\mathbf{x},t)-\mathbf{P}_\Psi(\mathbf{x},t)$. Therefore in the full 
Hilbert space we can define the operator
\begin{eqnarray}
\mathbf{E}^{(\pm)}(\mathbf{x},t)=(\mathcal{L}_t^{(\pm)}+\mathcal{M}_t^{(\pm)}+\mathcal{N}_t^{(\pm)})[\mathbf{E}(\mathbf{x},t)].
\end{eqnarray} 
Finally, from Eq.~\ref{machinchose} we deduce Eqs.~\ref{machinchoseplus} and \ref{machinchoseplusplus} which requires only $\mathcal{L}_t^{(\pm)}$.
%%%%%%%%%%%%%%
\section{Application of the operator $\mathcal{L}_t^{(\pm)}$ in the long time $t$ approximation }
 We start with the calculation of
\begin{eqnarray}
\boldsymbol{\Delta}_v^{(\pm)}(\tau,\mathbf{x},\mathbf{x'})=\int_{-\infty}^{+\infty} \frac{d\omega}{2\pi}\frac{\omega^2}{c^2}\mathcal{L}_\tau^{(\pm)}[e^{-i\omega\tau} \mathbf{G}_v(\mathbf{x},\mathbf{x'},\omega)]\label{greenplus1}\nonumber\\
\end{eqnarray}
In \cite{B} we showed that we have the dyadic expansion $\mathbf{G}_v(\mathbf{x},\mathbf{x'},\omega)=\mathbf{G}_{v,\bot}(\mathbf{x},\mathbf{x'},\omega)+\mathbf{G}_{v,||}(\mathbf{x},\mathbf{x'},\omega)$ 
 with for the transverse part \begin{eqnarray}
\mathbf{G}_{v,\bot}(\mathbf{x},\mathbf{x'},\omega)=\sum_{\alpha,j}\frac{c^2\Phi_\alpha(\mathbf{x})\Phi_\alpha^\ast(\mathbf{x'})\boldsymbol{\hat{\epsilon}}_{\alpha,j}\otimes\boldsymbol{\hat{\epsilon}}_{\alpha,j}}{\omega_\alpha^2-(\omega+i0^+)^2}\label{Gper}
\end{eqnarray}
and for the longitudinal part \begin{eqnarray}
\mathbf{G}_{v,||}(\mathbf{x},\mathbf{x'},\omega)
=\frac{-c^2\sum_{\alpha}\hat{\mathbf{k}}_\alpha\otimes\hat{\mathbf{k}}_\alpha\Phi_\alpha^\ast(\mathbf{x'})\Phi_\alpha(\mathbf{x})}{(\omega+i0^+)^2}\label{Gpar}
\end{eqnarray}
A direct application of $\mathcal{L}_t^{(\pm)}$ leads to 
\begin{eqnarray}
\mathcal{L}_\tau^{(\pm)}[e^{-i\omega\tau} \mathbf{G}_{v,\bot}(\mathbf{x},\mathbf{x'},\omega)]\nonumber\\=\sum_{\alpha,j}\frac{c^2\Phi_\alpha(\mathbf{x})\Phi_\alpha^\ast(\mathbf{x'})\boldsymbol{\hat{\epsilon}}_{\alpha,j}\otimes\boldsymbol{\hat{\epsilon}}_{\alpha,j}}{\omega_\alpha^2-(\omega+i0^+)^2}
\frac{1}{2}(1\pm\frac{\omega}{\omega_\alpha})e^{-i\omega\tau}\nonumber\\
\label{Gperbis}
\end{eqnarray} Moreover, if we use $\frac{1}{\omega_\alpha^2-(\omega+i0^+)^2}=\frac{1}{2\omega_\alpha}[\frac{1}{\omega_\alpha-\omega-i0^+}+\frac{1}{\omega_\alpha+\omega+i0^+}]$ and $\frac{1}{x-i0^+}=P.V.[\frac{1}{x}]+i\pi\delta(x)$ for $x$ real then we obtain
\begin{eqnarray}
\mathcal{L}_\tau^{(\pm)}[e^{-i\omega\tau} \mathbf{G}_{v,\bot}(\mathbf{x},\mathbf{x'},\omega)]\nonumber\\=\sum_{\alpha,j}\frac{c^2\Phi_\alpha(\mathbf{x})\Phi_\alpha^\ast(\mathbf{x'})\boldsymbol{\hat{\epsilon}}_{\alpha,j}\otimes\boldsymbol{\hat{\epsilon}}_{\alpha,j}}{2\omega_\alpha}
\frac{1}{2}(1\pm\frac{\omega}{\omega_\alpha})\nonumber\\ \times [i\pi\delta(\omega-\omega_\alpha)+P.V.[\frac{1}{\omega_\alpha-\omega}]\nonumber\\- i\pi\delta(\omega+\omega_\alpha)+P.V.[\frac{1}{\omega_\alpha+\omega}] ] e^{-i\omega\tau}
\label{Gper3} 
\end{eqnarray} which is actually equivalent to 
\begin{eqnarray}
\mathcal{L}_\tau^{(\pm)}[e^{-i\omega\tau} \mathbf{G}_{v,\bot}(\mathbf{x},\mathbf{x'},\omega)]\nonumber\\=\sum_{\alpha,j}\frac{c^2\Phi_\alpha(\mathbf{x})\Phi_\alpha^\ast(\mathbf{x'})\boldsymbol{\hat{\epsilon}}_{\alpha,j}\otimes\boldsymbol{\hat{\epsilon}}_{\alpha,j}}{2\omega_\alpha(\omega_\alpha\mp\omega\mp i0^+)} e^{-i\omega\tau}. \nonumber\\
\label{Gper4} 
\end{eqnarray}
Therefore, after integration in the complex plane  we get 
\begin{eqnarray}
\boldsymbol{\Delta}_v^{(+)}(t-t',\mathbf{x},\mathbf{x'})=\frac{i}{\hbar}\sum_{\alpha,j}\textbf{E}^{(v)}_{\alpha,j}(\mathbf{x})\otimes\textbf{E}^{(v)\ast}_{\alpha,j}(\mathbf{x}')
\nonumber\\ \times e^{-i\omega_\alpha(t-t')}\Theta(t-t').
\end{eqnarray}  and $\boldsymbol{\Delta}_v^{(-)}(t-t',\mathbf{x},\mathbf{x'})=(\boldsymbol{\Delta}_v^{(+)}(t-t',\mathbf{x},\mathbf{x'}))^\ast$. We emphasize that the longitudinal part  $\mathbf{G}_{v,||}(\mathbf{x},\mathbf{x'},\omega)$ doesn't contribute to $\boldsymbol{\Delta}_v^{(\pm)}(t-t',\mathbf{x},\mathbf{x'})$ as it can be shown directly.
Furthermore, if we take the imaginary part of Eqs.~\ref{Gper} and \ref{Gpar} and use once again the separation 
\begin{eqnarray}
\frac{1}{\omega_\alpha^2-(\omega+i0^+)^2}\nonumber\\=\frac{1}{2\omega_\alpha}[\frac{1}{\omega_\alpha-\omega-i0^+}+\frac{1}{\omega_\alpha+\omega+i0^+}]\nonumber\\
=\frac{i\pi}{2\omega_\alpha}(\delta(\omega-\omega_\alpha)-\delta(\omega+\omega_\alpha))\nonumber\\
+\frac{1}{2\omega_\alpha}(P.V.[\frac{1}{\omega_\alpha-\omega}])-P.V.[\frac{1}{\omega_\alpha+\omega}] )\nonumber\\
\end{eqnarray} we can directly demonstrate the rigorous equivalence  
\begin{eqnarray}
\boldsymbol{\Delta}_v^{(+)}(\tau,\mathbf{x},\mathbf{x'})=i\int_{0}^{+\infty} \frac{d\omega}{\pi}\frac{\omega^2}{c^2}\textrm{Imag}[\mathbf{G}_v(\mathbf{x},\mathbf{x'},\omega)]\nonumber\\ \times e^{-i\omega\tau} \theta(t-t')\label{greenplus2},
\end{eqnarray}  in which the contribution of  $\mathbf{G}_{v,||}(\mathbf{x},\mathbf{x'},\omega)$ vanishes  once again.  We emphasize that, despite some similarities, this result is different from Eq.~\ref{machinchoseplusplus} (in particular due to the presence of the imaginary part  and the Heaviside function $\theta(t-t')$ in Eq.~\ref{greenplus2}).
In order to evaluate  asymptotically  Eq.~\ref{greenplus2} for long time $t-t'$ we use the definition $\frac{\omega^2}{c^2}\mathbf{G}_{v}(\mathbf{x},\mathbf{x'},\omega)=\boldsymbol{\nabla}\times\boldsymbol{\nabla}\times[G_{v}(\mathbf{x},\mathbf{x'},\omega)\mathbf{I}]$ where $G_{v}(\mathbf{x},\mathbf{x'},\omega)=\frac{e^{i\omega/c|\mathbf{x}-\mathbf{x'}|}}{4\pi |\mathbf{x}-\mathbf{x'}|}$ is the standard scalar Green function of the Helmholtz equation in vacuum \cite{B}.
Now, we consider the integral 
\begin{eqnarray}
i\int_{0}^{+\infty} \frac{d\omega}{\pi}\textrm{Imag}[G_{v}(\mathbf{x},\mathbf{x'},\omega)]e^{-i\omega\tau}\nonumber\\ 
=\int_{0}^{+\infty} \frac{d\omega}{2\pi}\frac{e^{-i\omega(\tau-|\mathbf{x}-\mathbf{x'}|/c)}-e^{-i\omega(\tau+|\mathbf{x}-\mathbf{x'}|/c)}}{4\pi |\mathbf{x}-\mathbf{x'}|}.
\end{eqnarray} We remind that following Feynman~\cite{Feynman} we have  also the integral $\delta_+(x):=\int_0^{+\infty}\frac{d\omega}{\pi}e^{-i\omega x}=\frac{1}{i\pi}\frac{1}{x-i 0^+}=\frac{1}{i\pi}P.V. [1/x])+\delta (x)$ for  $x$ real. Therefore, we deduce
\begin{eqnarray}
i\int_{0}^{+\infty} \frac{d\omega}{\pi}\textrm{Imag}[G_{v}(\mathbf{x},\mathbf{x'},\omega)]e^{-i\omega\tau}\nonumber\\ 
=\frac{\delta_+(\tau-|\mathbf{x}-\mathbf{x'}|/c)-\delta_+(\tau+|\mathbf{x}-\mathbf{x'}|/c)}{8\pi |\mathbf{x}-\mathbf{x'}|}.
\end{eqnarray} We are interested in the regime $\tau\rightarrow + \infty$ and from its definition $\delta_+(\tau+|\mathbf{x}-\mathbf{x'}|/c)$ can be neglected. We thus  obtain 
 \begin{eqnarray}
i\int_{0}^{+\infty} \frac{d\omega}{\pi}\textrm{Imag}[G_{v}(\mathbf{x},\mathbf{x'},\omega)]e^{-i\omega\tau}\nonumber\\ 
\simeq \frac{\delta_+(\tau-|\mathbf{x}-\mathbf{x'}|/c)}{8\pi |\mathbf{x}-\mathbf{x'}|}\nonumber\\
=\int_{0}^{+\infty} \frac{d\omega}{2\pi} G_{v}(\mathbf{x},\mathbf{x'},\omega)e^{-i\omega\tau}.
\end{eqnarray} Finally, we deduce  the asymptotic result  
\begin{eqnarray}
\boldsymbol{\Delta}_v^{(+)}(\tau,\mathbf{x},\mathbf{x'})\simeq \int_{0}^{+\infty} \frac{d\omega}{2\pi}\frac{\omega^2}{c^2}\mathbf{G}_v(\mathbf{x},\mathbf{x'},\omega)e^{-i\omega\tau} \label{greenplus3},
\end{eqnarray}
which is valid in the limit $\tau\rightarrow + \infty$. In this formula  the absence of  a contribution like  $\int_{-\infty}^0 d\omega [...]$ clearly means that such a term is non resonant.
Moreover, from Eq.~\ref{formalquatre} and \ref{formalcinq} we know that generally speaking $\widetilde{\mathbf{E}}^{(0)}(\mathbf{x},\omega)$ and $\mathbf{G}(\mathbf{x},\mathbf{x'},\omega)$, i.e., the relevant fields for the inhomogeneous problem, can be calculated by using Lippman-Schwinger integrals which depend on the knowledge of the dyadic Green function in vacuum  $\mathbf{G}_v(\mathbf{x},\mathbf{x'},\omega)$.  Using the  previous asymptotic  of $\mathcal{L}_t^{(\pm)}[\mathbf{G}_v(\mathbf{x},\mathbf{x'},\omega)e^{-i\omega\tau}]$ it is  thus not difficult to deduce the generality of Eq.~\ref{machinchoseplusplus} in the long time limit $\tau\rightarrow + \infty$.   
%%%%%%%%%%%%%%%%%%%%%%%%

\end{document}